\documentclass[a4paper, 11pt]{article}
\usepackage[utf8]{inputenc}
\usepackage{lmodern}
\usepackage{amsmath}
\usepackage{fixmath}
\usepackage{exscale}
\usepackage{accents}
\usepackage{bigints}
\usepackage{physics}
\usepackage{commath}
\usepackage{float}
\usepackage{empheq}
\usepackage{subfigure}
\usepackage{caption}
\usepackage{physics}
\usepackage{graphicx}
\usepackage{upgreek}
\usepackage{mathtools}
\usepackage{textcomp}
\usepackage{amssymb}
\usepackage{bbold}
\usepackage{stackengine}
\usepackage{comment}
\usepackage{tabularx}
\usepackage[title]{appendix}

 \usepackage[top = 0.5in, left = 0.5 in, right = 0.5 in, bottom = 0.8 in]{geometry}

\usepackage[authoryear,round]{natbib}

\usepackage{bm}
\usepackage{authblk}

\numberwithin{equation}{section}
\usepackage[graphicx]{realboxes} 
\usepackage{rotating}

\usepackage{bbm}

\newcommand{\bsym}[1]{\boldsymbol{#1}}

\usepackage{setspace}
\usepackage[colorlinks=true, urlcolor=blue, pdfborder={0 0 0},citecolor=blue]{hyperref}

\title{An embedding-aware continuum thin shell formulation}
\date{}

\author[1]{Abhishek Ghosh}
\author[1]{Andrew McBride}
\author[2]{Zhaowei Liu}
\author[3]{Luca Heltai}
\author[1,4]{Paul Steinmann}
\author[1]{Prashant Saxena\thanks{Corresponding author email: \it{prashant.saxena@glasgow.ac.uk}}}
\affil[1]{\small James Watt School of Engineering, University of Glasgow, Glasgow G12 8LT, United Kingdom}
\affil[2]{\small College of Mechanics and Engineering Science, Hohai University, Nanjing, 211100, China}
\affil[3]{\small  University of Pisa,  Via Buonarroti 1/c
56127 Pisa (PI), Italy}
\affil[4]{\small Institute of Applied Mechanics, Friedrich-Alexander Universit\"at Erlangen-N\"urnberg, D-91052, Erlangen, Germany}

\begin{document}

\maketitle

\begin{abstract}
Cutting-edge smart materials are transforming the domains of soft robotics, actuators, and sensors by harnessing diverse non-mechanical stimuli, such as electric and magnetic fields.
Accurately modelling their physical behaviour necessitates an understanding of the complex interactions between the structural deformation and the fields in the surrounding medium.
For thin shell structures, this challenge is addressed by developing a shell model that effectively incorporates the three-dimensional field it is embedded in
by appropriately accounting for the relevant boundary conditions.
This study presents a  model for the nonlinear deformation of thin hyperelastic shells, incorporating Kirchhoff-Love assumptions and a rigorous variational approach.
The shell theory is derived from 3D nonlinear elasticity by dimension reduction while preserving the boundary conditions at the top and bottom surfaces of the shell.
Consequently, unlike classical shell theories, this approach can distinguish between pressure loads applied at the top and bottom surfaces, and delivers a platform to include multi-physics coupling.
Numerical examples are presented to illustrate the theory and provide a physical interpretation of the novel mechanical variables of the model.

\end{abstract}



\section{Introduction}
Advancements in smart materials based on multi-physics coupling are transforming the technological landscape in the field of soft robotics, actuators, and sensors. These cutting-edge materials harness non-mechanical stimuli from electric, magnetic, thermal, or chemical fields \citep{Mark_R_Jolly_1996, McKay_2010, kim-swell2012, Sheng2012}, presenting considerable challenges in modelling due to their complex physics. Specifically, the interplay between the surrounding space and the soft, deformable body is pivotal in electroelasticity, 
magnetoelasticity, and fluid-structure interaction scenarios \citep{dorfmann2014, Pelteret2020}.
These problems often involve thin structures, where one characteristic dimension is negligible compared to the others, ensuring an appreciable deformation response to external stimuli in the surrounding free space. 
Classically, thin structures are modelled as lower-dimensional manifolds embedded in three-dimensional space, incorporating appropriate kinematic simplifications. 
While these simplifications are often adequate to model such structures in isolation, they pose some challenges when coupling with the external ambient is required.
Consequently, there is a need for a thin shell formulation that is aware of the physical fields (electric/magnetic/pressure) in the embedding free-space. 
Working towards this aim, we present a novel thin shell theory derived from 3D nonlinear elasticity via dimension reduction that preserves the knowledge of fields and boundary conditions at the top and bottom surfaces.
This theory can be extended to modelling interactions between the thin shell and the free-space in electroelastic and magnetoelastic scenarios as has been preliminarily demonstrated in \citep{ghosh2023fullycoupled}.

Modelling slender structures, such as rods, membranes, plates, and shells, which exhibit both material and geometric nonlinearities is a particularly relevant due to their extensive use in engineering applications such as tyres, airbags, air springs, buffers, pneumatic actuators \citep{galley2019pneumatic}, and soft grippers \citep{hao2017modeling}.
When modelling physical phenomena on curved surfaces, definitions for geometric quantities (normal vectors, curvatures, etc.) and differential surface operators (gradients, divergence, etc.) are the key ingredients \citep{Paul_2015}. 
These quantities may be defined based on either two-dimensional, curvilinear local coordinates living on the manifold or on global coordinates of the surrounding, three-dimensional space.
A substantial body of literature exists on the  modelling of nonlinear membranes.
Using semi-analytical approaches and simplifications arising from axisymmetry, large inflation of hyperelastic membranes have been studied for circular  \citep{adkins1952large, hart1967large, yang1970axisymmetrical, SAXENA2019250}, cylindrical \citep{khayat1992inflation, guo2001large, pamplona2006finite, REDDY2018203}, spherical \citep{akkas1978dynamic, verron1999dynamic, XIE2016182}, and toroidal \citep{tamadapu2013finite, REDDY2017248,  pamulaparthi2019instabilities} membranes.

An overview of classical shell models with both analytical and computational approaches is provided, for example, in \citep{Simo1989OnSR, Cirak2000, Bischoff2004, kiendl2009isogeometric}, and more comprehensively in the textbooks \citep{Basar1985, Wempner2002, Blaauwendraad2014, Maria2017}. 
Various numerical techniques have been proposed to simulate nonlinear deformation of hyperelastic thin shells.
\citet{hughes2005isogeometric} pioneered the use of NURBS basis functions in finite elements which was applied to modelling thin shells by \citet{kiendl2009isogeometric}. 
\citet{kiendl2015isogeometric} further proposed an isogeometric thin shell formulation which was extended by \citet{liu2022computational} to model instabilities in large deformation of thin  shells.
Additionally, 
\citet{tepole2015isogeometric} and \citet{roohbakhshan2017efficient} developed isogeometric formulations for modelling thin soft biological tissues.
A complementary approach is based on the definition of a finite number of restriction operators from the full three-dimensional domain representing the slender structure and the lower dimensional manifold, together with their corresponding extension operators~\citep{HeltaiZunino-2023-a}. 
This approach uses the original (non-linear) model on the three-dimensional domain, while restricting the space of allowed solutions and test functions to a smaller subspace, isomorphic to a space on the lower dimensional manifold.
This effectively reduces the dimensionality of the problem, while naturally retaining the coupling with the surrounding ambient \citep{AlzettaHeltai-2020-a,HeltaiCaiazzoMueller-2021-a}.

In this work, a theory for modelling thin shells is derived from  three-dimensional incompressible hyperelasticity using the Kirchhoff--Love hypothesis \citep{Niordson1985}.
The effect of external loads is systematically captured in the formalism as the two-dimensional system is derived through suitable approximations.
This approach proves indispensible when analyzing a magnetoelastic shell, where careful consideration of changes in the surrounding magnetic field in the embedding space due to shell deformation is crucial \citep{ghosh2023fullycoupled}.
This contrasts with recent works on thin magnetoelastic structures that need to impose a ``weak magnetization'' assumption to simplify the problem by neglecting any perturbations in the magnetic field in the surrounding free space \citep{Barham_2008, REDDY2017248, Yan2020}.

The salient features of the present  work are the following:
\begin{enumerate}

\item \textbf{Distinction between top and bottom surface pressures:} Traditional shell models typically consider externally applied loads at the mid-surface of the structure. In contrast, our approach distinguishes between loads applied to the top and bottom surfaces, accurately capturing these boundary conditions. The implications of this deviation from conventional practice are explored through numerical examples.


\item \textbf{Comprehensive characterisation of nonlinear deformation:} 
Nonlinear deformation of a thin Kirchhoff--Love shell is completely characterised by the deformed mid-surface position vector $(\bsym{r})$, thickness stretch $(\lambda)$, and the unit normal vector $(\bsym{n})$ to the deformed surface.
Variational formulations of conventional shell models are directly based on $\bsym{r}$ and therefore only need to consider perturbations $\delta \bsym{r}$ in the analysis.
Our approach, derived systematically from three to two dimensions, naturally incorporates perturbations   $\delta \lambda$ and $\delta \bsym{n}$.
This enrichment adds complexity to the derivation of the shell system of equations and necessitates a unique application of Green’s theorem. 
This study addresses these complexities, providing a generalised system of partial differential equations with boundary terms that encompass these effects.



\item \textbf{Surface-based partial differential equations:} The partial differential equations governing shell equilibrium derived in this work are entirely based on the surface fields.
Series expansion of all fields along the thickness coordinate ensures all information from the top and bottom surfaces is captured and an accuracy up to the linear order of the through-thickness parameter is maintained.
This is in contrast to conventional approaches \citep{Cirak:2001aa, kiendl2015isogeometric} where stress resultants are integrated through the thickness to formulate a virtual work principle based on effective normal force and bending moment.

\end{enumerate}

\subsection{Outline}
The structure of this paper is as follows:
The preliminaries of nonlinear elasticity are presented in Section \ref{sec: elasticity}.
Section \ref{sec: KL thin shell} presents the necessary geometric definitions and the kinematics of the Kirchhoff--Love thin shell formulation.
Through a variational approach, equations of 3D elasticity are systematically reduced to the mid-surface of the shell in Section \ref{sec: shell eqns from 3D}.
Section \ref{sec: numerical} provides the solution of two example boundary value problems  to illustrate the novel features of the derived shell equations.
Section \ref{sec: conclusions} concludes with closing remarks and an  outline for future research in this area.
Equations corresponding to the geometrical description of the shell are presented in Appendix A, details about the Green's theorem are presented in Appendix B, and the equations corresponding to the variation of relevant physical variables are presented in Appendix C.

\subsection{Notation}
A variable typeset in a normal weight font represents a scalar. A bold weight font denotes a first or second-order tensor. 
Latin indices $i,j,k,\dots$ vary from $1$ to $3$ while Greek indices $\alpha, \beta, \gamma,\dots$, used for surface variable components, vary from $1$ to $2$. Einstein summation convention is used throughout.
$ \mathbf{e}_i$ represent the basis vectors of an orthonormal $(x_1, x_2, x_3)$ coordinate system in three-dimensional Euclidean space. 
The three covariant basis vectors for a surface point are denoted as $\mathbf a_i$, where $\mathbf a_\alpha$ are the two tangential vectors and $\mathbf a_3$ as the normal vector with
$\theta^\alpha$ and $\eta$ as the respective coordinates. 

The comma symbol in a subscript represents partial derivative with respect to the surface coordinates, for example, $A_{,\beta}$ is the partial derivative of $A$ with respect to  $\theta^\beta$. 
The scalar product of two vectors $\bm{p}$ and $\bm{q}$ is denoted $\bm{p} \cdot \bm{q}$, and the dyadic product of these vectors is a second order tensor $\bm{H}=\bm{p}\otimes\bm{q}$.
Operation of a second order tensor $\bm{H}$  on a vector $\bm{p}$ is given by $\bm{H}\bm{p}$. The scalar product of two tensors $\bm{H}$ and $\bm{G}$ is denoted $\bm{H} : \bm{G}$. 
The notation $\norm{\cdot}$ represents the Euclidean norm. 
For a second order tensor in its component form $\bm{H}=H^{ij}\mathbf a_i\otimes\mathbf a_j$, the matrix is denoted $\left[H^{ij}\right]$.
Circular brackets $( \,)$ are used to denote the parameters of a function and square brackets $[ \,]$ are used to group mathematical expressions.  
\section{Non-linear elastostatics preliminaries}
\label{sec: elasticity}
Consider a homogenous body occupying regions  of space ${\mathcal{B}}_{\mathrm{0}}$ and $\mathcal{B}$ in ${\mathbb{R}}^3$ in its reference and deformed configuration, respectively, as depicted in Figure 1.
The boundaries of the regions are $\partial{\mathcal{B}}_{\mathrm{0}}$ and $\partial\mathcal{B}$. 
A point $\bm{X} \in \mathcal{B}_0$ is 
connected to a point ${\bm{x}} \in {\mathcal{B}}$ through a bijective mapping
${\bm{\chi}} : {\mathcal{B}}_{\mathrm{0}} \rightarrow \mathcal{B}$.
The deformation gradient $\bm{F}$ is defined as $
\bm{F}
=\dfrac{\partial \bm{\chi}}{\partial \bm{X}}
$ with the Jacobian, $J=\mathrm{det}\bm{F} > 0$. 
 The right  Cauchy-Green  tensor is then defined by $
\bm{C}={\bm{F}}^{\mathrm{T}}\bm{F}$.

The hyperelastic material model assumes that the 
free energy density function per unit reference volume is of the form
$
\Omega=\Omega \left( \bm{F} \right).
$
Objectivity and isotropy require that the free energy  takes the form 
\begin{equation}
\Omega = \breve{{\Omega}}\left(\bm{C}\right)=\widetilde{{\Omega}}\left(I_1,I_2,I_3\right),
\label{eq:free_energy2}
\end{equation}
where $I_1, I_2, I_3$ are scalar invariants of $\bm{C}$, given by
\begin{equation}
I_1 = \mathrm{tr} \bm{C}, \quad
I_2 = \dfrac{1}{2}\left[{\left[\mathrm{tr}\bm{C}\right]}^{2}-\mathrm{tr}{\bm{C}^{2}}\right],  \quad \text{and} \quad
I_3 = \mathrm{det}\bm{C} = J^2.
\end{equation}
 Under the constraint of incompressibility ($J\equiv 1$), the energy density function can be further simplified to
\begin{equation}
    \widetilde{{\Omega}}\left(I_1,I_2,I_3\right)=\grave{{\Omega}}\left(I_1,I_2\right).
\end{equation}
For incompressible hyperelastic solids, 
 the  Piola  stress tensor is given as
\begin{equation}
\bm{P}= \dfrac{\partial {\Omega}}{\partial\bm{F}} - p{\bm{F}}^{-\mathrm{T}},
\label{eq:cauchy_constitutive_general}
\end{equation}
 where $p$ is a  Lagrange multiplier due to the incompressibility constraint and it is identified as the hydrostatic pressure.

\section{Kirchhoff--Love thin shell}
\label{sec: KL thin shell}

In this section, we present the necessary relations to model thin shells using the classical Kirchhoff-Love assumption \cite[see, e.g.][]{Niordson1985}. 



\subsection{Shell geometry and kinematics }
\begin{figure}
  \begin{center}
  \def\svgwidth{12cm} 
  \input{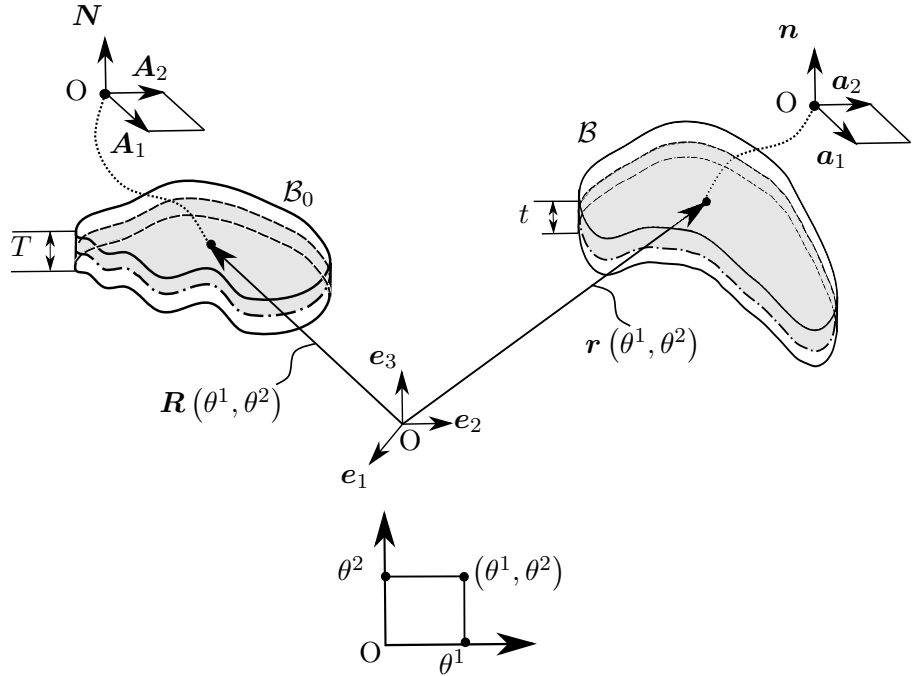}
    \caption{The position vectors at the mid-surface of a shell in the two configurations along with the local triads. The  parametric coordinates $\{\theta^\alpha\}$ are also shown.}
      \label{fig:metrics}
  \end{center}
\end{figure}

 Consider the three-dimensional  body introduced in Section \ref{sec: elasticity} to be a thin shell as shown in Figure \ref{fig:metrics}. 
 Each point $\bm{X} \in {\mathcal{B}}_{\mathrm{0}}$ is mapped from the parametric domain defined by the coordinate system $\{ \theta^1, \theta^2, \eta\}$.
 The Kirchhoff-Love hypothesis states that for thin shell structures, lines perpendicular to the mid-surface of the shell remain straight and perpendicular to the mid-surface after deformation.
 Hence, assuming the shell has a thickness $T\left(\theta^\alpha\right)$ in the reference configuration, the point $\bm{X}$  can be defined using a point on the mid-surface ${S}_{\mathrm{m}}$ of the shell, ${\bm{R}} \in {S}_{\mathrm{m}}$, and the associated unit normal vector ${\bm{N}}$ by
 \begin{equation}
\bm{X} = {\bm{R}} + \eta {\bm{N}},
\label{eq: r x n relation}
 \end{equation}
 where $\eta \in [-T/{2}, T/{2}]$. 
The same point on the mid-surface in the deformed configuration  $\bm{r}$ is related to $\bm{R}$ by 
\begin{equation}
\bm{r} = \bm{R} + \bm{u},
\label{eq:reference_to_deformed}
\end{equation}
where $\bm{u}$ denotes the mid-surface displacement vector. 
In a similar fashion, each point $\bm{x} = \bsym{\chi} (\bsym{X}) \in \mathcal{B}$ can be expressed as 
\begin{equation}
\bm{x}=\bm{r}+\eta \bm{d}, 
\label{eq:def_map_body}
\end{equation} 
where the director $\bm{d}=\lambda \bm{n}$, $\bm{n}$ being the unit normal vector in the deformed configuration and $\lambda$  the through-thickness stretch defined by
\begin{equation}
\lambda= \frac{t}{T}.
\end{equation}
Here $t(\theta^{\alpha})$ is the shell thickness after deformation. 
The ensuing analysis is simplified by employing the long-wave assumption for thin structures ($\lambda_{,\alpha} \approx 0$) that states that $\lambda$ changes slowly along the mid-surface of the shell \citep{ kiendl2015isogeometric, liu2022computational}.
This assumption helps simplify the mathematical derivations in obtaining the shell system of equations, since the shifter relations in the deformed configuration no longer contain a  term involving the parametric derivative of $\lambda$.
Mathematical details on this are provided in Appendix \ref{app: natural basis}.
For thin shells modelled with the Kirchhoff-Love assumption, transverse shear and localised necking are unlikely to occur and therefore the long-wave assumption is acceptable.
Indeed, 
the long-wave assumption works well for most problems concerning nonlinear deformation of thin shells modelled using the Kirchhoff-Love model.
However, it is recognised that it should be discarded for problems that require modelling of localised necking and transverse shear deformation. 


 Table \ref{demo-table1} lists the  surface parameters  and Table \ref{demo-table2} presents the surface and volume elements of the shell. 
 The expressions and  associated derivations are elaborated on in Appendix \ref{app: KL shell geometry}.
The boundaries, $\partial{\mathcal{B}}_{\mathrm{0}}$ and $\partial \mathcal{B}$ can be written as
$ \partial{\mathcal{B}}_{\mathrm{0}}= S_\mathrm{t} \cup   S_\mathrm{b}  \cup   S_\mathrm{l}$ and $\partial \mathcal{B}= s_\mathrm{t} \cup  s_\mathrm{b} \cup  s_\mathrm{l}$,
where the subscripts, t, b, and l, represent the top, bottom, and  lateral surfaces, respectively. Here, the top surface is  the boundary that is reached along the unit outward normal vector.

\begin{sidewaystable}
\small
\centering
\label{tab: 1}
\caption{\label{demo-table1} A list of parameters used to describe the geometry of the thin shell}
 \begin{tabular}{||l l l||} 
 \hline
 Surface Parameters & Reference Configuration & Deformed Configuration \\ [0.4ex] 
 \hline\hline
 Covariant basis vectors at the mid-surface  & $\bm{A}_\alpha$ & $\bm{a}_\alpha$ \\
 Covariant metric tensor at the mid-surface &$ A_{\alpha \beta}$ & 
 $a_{\alpha \beta}$ \\  
  Determinent of the covariant metric tensor at the mid-surface &$ A $ & 
 $a$ \\ 
 Stretch at the mid-surface & -- -- & $\hat{a} =a/A$ \\
 Contrvariant basis vectors at a mid-surface  & $\bm{A}^\alpha$ & $\bm{a}^\alpha$ \\
 Contravariant metric tensor at the mid-surface & $A^{\alpha \beta}$ & $a^{\alpha \beta}$  \\
  Determinent of the contravariant metric tensor at the mid-surface & $A^{-1}$ & $a^{-1}$  \\
 Christoffel symbol at the mid-surface & ${\Gamma}^\alpha_{\beta \gamma}$ & ${\gamma}^\alpha_{\beta \gamma}$\\
 Parametric derivative of the metric at the mid-surface & ${A}_{,\zeta}
 $ & ${a}_{,\zeta}
 $ \\
  Normal  at the mid-surface & $\bm{N}$ & $\bm{n}$ \\
 Tangent  on the bounding curve of the  mid-surface  & $\bm{\tau}$ & $--$ \\
 In-plane  normal on the bounding curve of the  mid-surface & $\bm{\nu} = \nu_\alpha \bm{A}^\alpha$ & $\overline{\bsym{\nu}} = \nu_\alpha \bsym{a}^{\alpha}$ \\
  Projection tensor at the mid-surface & $\bm{I}$ & $\bm{i}$ \\
 Curvature tensor at the mid-surface & $\bm{K}$ & $\bm{\kappa}$ \\
 Mean curvature at the mid-surface & $H$ & $h$ \\
 Gaussian curvature at the mid-surface & $K$ & $\kappa$ \\
 Covariant basis vectors at a shell-point & $\bm{G}_\alpha=\bm{M} \bm{A}_\alpha \ \text{and} \ \bm{M}=\bm{I}-\eta \bm{K}$ & $\bm{g}_\alpha=\bm{\mu} \bm{a}_\alpha \ \text{and} \ \bm{\mu}=\bm{i}-\eta \lambda\bm{\kappa}$  \\
 Covariant metric tensor at a shell-point & $G_{\alpha \beta}$ & $g_{\alpha \beta}$\\
 Contravariant basis vectors at a shell-point & ${\bm{G}}^{\alpha}={\bm{M}}^{-\mathrm{T}} {\bm{A}}^{\alpha}$ & ${\bm{g}}^{\alpha}={\bm{\mu}}^{-\mathrm{T}} {\bm{a}}^{\alpha}$  \\
 Contravariant metric tensor at a shell-point & $G^{\alpha \beta}$  & $g^{\alpha \beta}$   \\
 Tangent  at a shell-point on the lateral surface & $\bm{\tau}_\mathrm{l}=\bm{M}\bm{\tau}/ c \ \text{and} \ c=\norm{\bm{M}\bm{\tau}}$ & $--$ \\
 In-plane normal  at a shell-point on the lateral surface & $\bm{\nu}_\mathrm{l}={c}^{-1}\left[\bm{I}+\eta\left[\bm{K}-2H\bm{I}\right]\right]\bm{\nu}$ & $--$ \\
  [1ex] 
 \hline
 \end{tabular}
\end{sidewaystable}

\begin{table}
\centering
\caption{\label{demo-table2} Surface and volume elemental quantities}
 \begin{tabular}{||l l l||} 
 \hline
 Volume/Surface elements  & Reference Configuration & Deformed Configuration  \\ [0.5ex] 
 \hline\hline
 Area for the convected coordinates & $dP$ & $dP$ \\
 Elemental area at the curved mid-surface & $dS_\mathrm{m}$ & $ds_\mathrm{m}$ \\ 
 Area at a shell-point & $dS=M d{S}_\mathrm{m}, \,  M=\mathrm{det} \bm{M}$ & $ds=\mu {\hat{a}}^{1/2} dS_\mathrm{m}, \ \mu=\mathrm{det} \bm{\mu}$ \\
 Area at the top surface & $dS_\mathrm{t}$  & $ds_\mathrm{t}$  \\
 Area at the bottom surface & $dS_\mathrm{b}$  & $ds_\mathrm{b}$  \\
 Area on the lateral surface & $dS_\mathrm{l}=c \, dl \, d\eta$  & $--$  \\
  Volume at a shell-point & $dV=dS \, d\eta$ & $--$ \\
 [1ex] 
 \hline
 \end{tabular}
 
\end{table}



The deformation gradient 
can be expanded so as to separate the thickness variable $\eta$ from the surface parameters. 
\begin{eqnarray}
\bm{F}
= {F}^{ \alpha}_{0_\beta}\ \bm{a}_{\alpha} \otimes {\bm{A}}^{\beta}+\eta {F}^{  \alpha}_{1_\beta}\ \bm{a}_{\alpha} \otimes {\bm{A}}^{\beta} + \eta^2 {F}^{ \alpha}_{2_\beta}\ \bm{a}_{\alpha} \otimes {\bm{A}}^{\beta} + \bm{\mathcal{O}}(\eta^3) +\lambda \bm{n}\otimes\bm{N},
\label{def_gradient}
\end{eqnarray}
where the individual scalar components are written using the geometric parameters provided in Table \ref{demo-table1} as
\begin{eqnarray}
{F}^{ \alpha}_{0_\beta}=\delta^\alpha_\beta,\quad
{F}^{ \alpha}_{1_\beta}=-\lambda {b}_\beta^{ \ \alpha}+{B}_\beta^{ \ \alpha}, \quad \text{and} \quad 
{F}^{  \alpha}_{2_\beta}={B}_\delta^{ \ \alpha}{B}_\beta^{ \ \delta}-\lambda {b}_\delta^{ \ \alpha} {B}_\beta^{ \ \delta} .
\end{eqnarray}
Here  ${B}_\beta^{ \ \alpha}$ and ${b}_\beta^{ \ \alpha}$ are the components of the curvature tensors $\bm{K}$ and $\bm{\kappa}$, respectively.
Applying the same decomposition to the inverse of the deformation gradient tensor, we obtain
\begin{eqnarray}
\bm{F}^{-1}
={F}^{-1 \alpha}_{0_\beta}\ {\bm{A}}_{\alpha} \otimes \bm{a}^{\beta}
+
\eta {F}^{-1  \alpha}_{1_\beta}\ {\bm{A}}_{\alpha} \otimes \bm{a}^{\beta} 
+ 
\eta^2 {F}^{-1  \alpha}_{2_\beta}\ {\bm{A}}_{\alpha} \otimes \bm{a}^{\beta} + \bm{\mathcal{O}}(\eta^3) +\frac{1}{\lambda} \bm{N}\otimes \bm{n},
\label{inv_def_gradient}
\end{eqnarray}
with
\begin{equation}
{F}^{-1 \alpha}_{0_\beta}=\delta^\alpha_\beta,\quad
{F}^{-1  \alpha}_{1_\beta}=\lambda {b}_\beta^{ \ \alpha}-{B}_\beta^{ \ \alpha}, \quad \text{and} \quad
{F}^{-1  \alpha}_{2_\beta}=\lambda^2{b}_\delta^{ \ \alpha}{b}_\beta^{ \ \delta}-\lambda {B}_\delta^{ \ \alpha} {b}_\beta^{ \ \delta}.
\end{equation}


Upon recognising that $\bm{a}_\alpha \cdot \bm{n}=0$, the right Cauchy--Green  tensor $\bm{C} $ can be expanded as
\begin{equation}
\bm{C}
={C}_{0_{\alpha\beta}} {\bm{A}}^{\alpha}\otimes {\bm{A}}^{\beta} + 
\eta {C}_{1_{\alpha\beta}}{\bm{A}}^{\alpha}\otimes {\bm{A}}^{\beta} + 
\eta^2 {C}_{2_{\alpha\beta}} {\bm{A}}^{\alpha}\otimes {\bm{A}}^{\beta} + \bm{\mathcal{O}}(\eta^3)+ \lambda^2 \bm{N}\otimes \bm{N},
\end{equation}
where the individual components are 
\begin{eqnarray}
{C}_{0_{\alpha\beta}}&=&a_{\alpha \beta},\nonumber \\
{C}_{1_{\alpha\beta}}&=&-\lambda {b}_\beta^{ \ \gamma}a_{\gamma \alpha}-\lambda {b}_\alpha^{ \ \gamma}a_{\gamma \beta} +{B}_\alpha^{ \ \gamma}a_{\gamma \beta}+{B}_\beta^{ \ \gamma}a_{\gamma \alpha}, \nonumber\\
{C}_{2_{\alpha\beta}}&=&{B}_\alpha^{ \ \gamma}{B}_\beta^{ \ \delta}a_{\delta \gamma }
+\lambda^2 {b}_\alpha^{ \ \delta}{b}_\beta^{ \ \gamma}a_{\delta \gamma }
+{B}_\gamma^{ \ \delta}{B}_\alpha^{ \ \gamma}a_{\delta \beta }  
+{B}_\gamma^{ \ \delta}{B}_\beta^{ \ \gamma}a_{\delta \alpha }
-\lambda {b}_\gamma^{ \ \delta} {B}_\beta^{ \ \gamma}a_{\delta \alpha } 
-\lambda {b}_\gamma^{ \ \delta} {B}_\alpha^{ \ \gamma}a_{\delta \beta } 
\nonumber \\
&& - \lambda {B}_\beta^{ \ \gamma}{b}_\alpha^{ \ \delta} a_{\delta \gamma } 
-\lambda {B}_\alpha^{ \ \gamma}{b}_\beta^{ \ \delta} a_{ \gamma \delta}.
\end{eqnarray}

The decompositions in equations allow us to write the relevant quantities at the mid-surface for later use. 
The mid-surface right Cauchy-Green tensor $\bm{C}_\mathrm{m}$  is given by
\begin{equation}
    \bm{C}_\mathrm{m}=\bm{C}\Big|_{\eta=0}={C}_{0_{\alpha\beta}} {\bm{A}}^{\alpha}\otimes {\bm{A}}^{\beta} +\lambda^2 \bm{N}\otimes \bm{N}.
\end{equation}
Its determinant squared is expressed as
\begin{equation}
    J_0^2 = \mathrm{det}\bm{C}_\mathrm{m}=
    \mathrm{det}\left[{C_{\mathrm{m}_j}}^{  i}\right]=
    \mathrm{det}\left[{C}_{\mathrm{m}_{ j k}}{A}^{ki }\right]
    =\frac{ \mathrm{det}\left[{C}_{\mathrm{m}_{ ij}}\right]}{\mathrm{det}\left[{A}_{ij }\right]}=\frac{ \mathrm{det}\left[a_{\alpha \beta}\right] \lambda^2}{\mathrm{det}\left[{A}_{\alpha \beta}\right]}
    =\frac{ a \lambda^2}{A},
\end{equation}
where ${A}_{i j} = {\bm{A}}_{i} \cdot {\bm{A}}_{j}$ and ${A}^{i j} = {\bm{A}}^{i} \cdot {\bm{A}}^{j}$ are the components of the three-dimensional covariant and contravariant metric tensors , respectively, on the mid-surface  with $ {\bm{A}}_{3}= 
    {\bm{A}}^{3}
    ={\bm{N}}$.
The incompressibility constraint
$J_0=1$
implies
\begin{equation}
    \lambda={\hat{a}}^{-1/2} ,
    \label{eq:through_thickness}
\end{equation}
where $\hat{a}=a/A$.


\subsection{Divergence of stress and Green's theorem}


The divergence of the Piola stress tensor is required for derivation of the governing equations for the Kirchhoff-Love thin shell in later sections.
It is given by
\begin{eqnarray}
   \mathrm{Div}\bm{P}
   &=&\bm{P}_{, \alpha}  {\bm{G}}^\alpha + \frac{\partial \bm{P}}{\partial \eta}  {\bm{N}}, \nonumber \\
     &=&\bm{A}_0+\eta \bm{A}_1 + \eta^2 \bm{A}_2 +\bm{\mathcal{O}}(\eta^3),
  \label{eq:div_total_stress}
\end{eqnarray}
where ${\bm{G}}^{\alpha}={\bm{M}}^{-\mathrm{T}} {\bm{A}}^{\alpha} + \bm{\mathcal{O}}(\eta^2)$. 
$\bm{A}_0, \bm{A}_1, \bm{A}_2$ are the zeroth; first; and second; order terms independent of the thickness variable $\eta$ and are given by
\begin{eqnarray}
  \bm{A}_0&=&{\bm{P}_0}_{, \alpha}  {\bm{A}}^\alpha + \bm{P}_1  \bm{N}, \nonumber \\
  \bm{A}_1&=&{B}_\delta^{ \ \alpha} {\bm{P}_0}_{, \alpha}  {\bm{A}}^\delta+{\bm{P}_1}_{, \alpha} {\bm{A}}^\alpha
  +\bm{P}_2 \bm{N}, \nonumber \\
  \bm{A}_2 &=& {B}_\zeta^{ \ \alpha}{B}_\delta^{ \ \zeta}{\bm{P}_0}_{, \alpha}  {\bm{A}}^\delta + {B}_\delta^{ \ \alpha} {\bm{P}_1}_{, \alpha}  {\bm{A}}^\delta + \frac{1}{2}{\bm{P}_2}_{, \alpha} {\bm{A}}^\alpha 
 + \frac{1}{2}\bm{P}_3  \bm{N},
\end{eqnarray}
with
\begin{eqnarray}
    \bm{P}
    &=&\bm{P}_0+\eta \bm{P}_1 + \frac{\eta^2}{2}\bm{P}_2 + \frac{\eta^3}{6}\bm{P}_3+\bm{\mathcal{O}}(\eta^4) .
    \label{eq:total_stress_taylor}
\end{eqnarray}
The  stress  at the top and bottom boundaries can be obtained by setting  $\eta= T/2$ and $-T/2$, respectively.
We will require the zeroth. and the first. order terms of the Piola stress explicitly given as
\begin{equation}
    \bm{P}_0=\bm{\mathcal{P}}_0-p_0\bm{F}^{-\mathrm{T}}_0,\quad \text{and} \quad
    \bm{P}_1=\bm{\mathcal{P}}_1-p_0\bm{F}^{-\mathrm{T}}_1-p_1\bm{F}^{-\mathrm{T}}_0,
    \label{eq:Lag_comp}
\end{equation}
with 
\begin{equation}
  \frac{\partial \Omega}{\partial\bm{F}} = 
  \bm{\mathcal{P}}(\eta, \theta^\alpha)
  =\bm{\mathcal{P}}_0+\eta\bm{\mathcal{P}}_1+\bm{\mathcal{O}}(\eta^2),\quad \text{and} \quad \bm{F^{-\mathrm{T}}}=\bm{F^{-\mathrm{T}}}_0+\eta \bm{F^{-\mathrm{T}}}_1+\bm{\mathcal{O}}(\eta^2). 
\end{equation}

Mathematical analysis for shell mechanics often requires change of integration from the parametric domain to the shell boundary with expressions that involve parametric derivatives (see, for example, Section 4.2). 
This requires a modified form of Green's theorem which states that for a vector with components ${T}^\alpha$ the following relation holds
\begin{equation}
\int\limits_{ P}\left[{A}^{1/2}{T}^\alpha\right]_{, \alpha}dP=\int\limits_{\mathcal{C}_\mathrm{m}}{T}^\alpha {\nu}_\alpha dl.
   \label{eq:dom_to_bound}
\end{equation}
Here $\mathcal{C}_\mathrm{m}$ is the boundary of the curved mid-surface $S_\mathrm{m}$ .
A complete derivation of this result is provided in Appendix \ref{app: Green}.

\section{Derivation of shell equations from 3D elasticity}
\label{sec: shell eqns from 3D}
We now systematically derive the necessary shell equations starting from a classical 3D potential energy functional for a hyperelastic body.

\subsection{Variational formulation in three dimensions}
The total potential energy of an incompressible elastic body undergoing a deformation $\bm{\chi}$ is given by
\begin{eqnarray}
     \Pi[\bm{\chi},  p] = \int\limits_{{\mathcal{B}}_0}\Omega\left( \bm{F} \right)dV-
    \int\limits_{\mathcal{B}_0}p[J-1]dV
    -\int\limits_{\mathcal{B}_0}\bm{\mathfrak{B}}\cdot \bm{\chi}dV -\int\limits_{ {\mathcal{C}}_\mathrm{m} \setminus {\mathcal{C}}^{\mathrm{u}}_{\mathrm{m}}}
    \bm{t}_\ell \cdot\bm{\chi}dl
    -\int\limits_{ s_\mathrm{t}}\bm{p}_\mathrm{t}\cdot \bm{\chi}ds_\mathrm{t}
     -\int\limits_{ s_\mathrm{b}}\bm{p}_\mathrm{b} \cdot \bm{\chi}ds_\mathrm{b}, 
     \label{eq:energy}
\end{eqnarray}
where $\left(\bm{\chi},  p\right)$ is the  set of generalised solution variables.
The body force field per unit volume is $\bm{\mathfrak{B}}$ and $\bm{t}_\ell$ is the applied traction due to dead load at  ${C}_\mathrm{m}$.
The Dirichlet part of the boundary is denoted by ${C}_\mathrm{m}^\mathrm{u}$.
We defined ${p}_\mathrm{t}\left(\theta^\alpha\right)$ and ${p}_\mathrm{b}\left(\theta^\alpha\right)$ as the magnitudes of external pressure at the top and bottom surfaces of the shell, respectively, such that 
\begin{equation}
    \bm{p}_\mathrm{t}=-p_\mathrm{t}\bm{n}_\mathrm{t}, \quad \text{and} \quad \bm{p}_\mathrm{b}=-p_\mathrm{b}\bm{n}_\mathrm{b},
    \label{eq:press_term_forms}
\end{equation}
with $ \bm{n}_\mathrm{t} = \bm{n}$ and $ \bm{n}_\mathrm{b} = - \bm{n}$ as the outward unit normals at these two surfaces  in the current configuration.
Upon arbitrary variations $\delta\bm{\chi}$ and $\delta p$ of the solution variables,  the first variation of the energy functional becomes
\begin{eqnarray}
    \delta \Pi[\bm{\chi},  p; \delta \bm{\chi},  \delta p]&=& \underbrace{\delta \left[\int\limits_{{\mathcal{B}}_0}\Omega\left( \bm{F} \right)dV\right]
    -\int\limits_{{\mathcal{B}}_0}p\delta JdV}_I
    -\int\limits_{{\mathcal{B}}_0}\delta p [J-1]dV 
    \nonumber \\
    && -\int\limits_{{\mathcal{B}}_0}\bm{\mathfrak{B}}\cdot \delta\bm{\chi}dV 
    - \int\limits_{ {\mathcal{C}}_\mathrm{m} \setminus {\mathcal{C}}_\mathrm{m}^\mathrm{u}}\bm{t}_\ell \cdot\delta\bm{\chi}dl
    -\int\limits_{ s_\mathrm{t}}\bm{p}_\mathrm{t}\cdot \delta\bm{\chi}ds_\mathrm{t}
     -\int\limits_{ s_\mathrm{b}}\bm{p}_\mathrm{b}\cdot  \delta\bm{\chi}ds_\mathrm{b}. 
     \label{eq:var_2}
\end{eqnarray}
Upon using Equation \eqref{eq:cauchy_constitutive_general} and an application of the divergence theorem, the first and second terms of the above Equation \eqref{eq:var_2} can be written as
\begin{align}
I= - \int\limits_{{\mathcal{B}}_0}\mathrm{Div}\bm{P}\cdot\delta\bm{\chi}dV
 + \int\limits_{ S_\mathrm{t}}\bm{P}{\bm{N}}_\mathrm{t}\cdot\delta\bm{\chi}dS_\mathrm{t}
 +\int\limits_{ S_\mathrm{b}}\bm{P}{\bm{N}}_\mathrm{b}\cdot\delta\bm{\chi}dS_\mathrm{b} 
 +\int\limits_{ S_\mathrm{l}}\bm{P}{\bm{\nu}}_\mathrm{l} \cdot\delta\bm{\chi}dS_\mathrm{l} 
,
 \label{eq:var_inside_body}
\end{align}
The terms involving the Piola stress tensor are addressed further with in Section \ref{sec: piola terms}. 
Note that, ${\bm{N}}_\mathrm{t} = \bm{N}$ and ${\bm{N}}_\mathrm{b} = - \bm{N}$  are the  unit outward normals at the top and bottom surfaces of the shell in the reference configuration, respectively.
The remaining terms in Equation  \eqref{eq:var_2}, that is, the virtual work done by the dead load traction, body force and pressures are addressed in Section \ref{sec: external loads terms}, where their contributions to a modified variational form for a Kirchoff-Love thin shell  are discussed.

\subsection{Derivation of the shell model}
The subsequent discussion introduces the key steps in deriving the Kirchhoff-Love shell equations. It will shown that   the derived equations can achieve an accuracy of $\mathcal{O}\left(T\right)$.
The  solution variables are alternatively expressed as quantities defined over the mid-surface of the shell, $\bm{r}$ and $ p_0$, such that
\begin{equation}
\delta \Pi[\bm{\chi},  p; \delta \bm{\chi}, \delta p]=\delta \widetilde{\Pi}[\bm{r},p_0;  \delta \bm{r}, \delta p_0].
\end{equation}
In the following sections, all  integrals in Equation \eqref{eq:var_2} are systematically converted from a dependence on $(\bm{\chi},  p)$ to a dependence on $(\bm{r},p_0)$.
This allows for the determination of the governing equations over the mid-surface of the shell. 


\textbf{Remark:} 
Shell models are usually simplified by incorporating the plane-stress assumption due to a shell being a thin structure.
In the case of shells made of incompressible hyperelastic materials, application of the plane-stress condition leads to  the explicit determination of the Lagrange multiplier as a function of deformation variables \citep{kiendl2015isogeometric, liu2022computational}.
Consequently, the Lagrange multiplier within the equations changes in nature from being an independent field to a constitutive expression.
In the subsequent derivations, the Lagrange multiplier is maintained as an independent field until the plane-stress condition is applied in Section \ref{sec: strong form}.



\subsubsection{Terms involving the Piola stress tensor ($I$)}
\label{sec: piola terms}

Upon using  Equations \eqref{diff_vol} and \eqref{eq:undeformed_area_arb}, where the latter relates the undeformed elemental area at a shell-point to the undeformed elemental area on the midsurface, the first term of $I$ in Equation \eqref{eq:var_inside_body} can be written as

\begin{align}
-\int\limits_{{\mathcal{B}}_0}\mathrm{Div}\bm{P}\cdot\delta\bm{\chi}dV &= -\int\limits_{S}\int\limits_{\eta}\mathrm{Div}\bm{P}\cdot\delta\bm{\chi} d\eta dS = -\int\limits_{ S_\mathrm{m}}\int\limits_{\eta}\mathrm{Div}\bm{P}\cdot \delta\bm{\chi} Md\eta d{\mathcal{S}}_\mathrm{m}, \\
& \quad \text{substituting} \, \bsym{\chi} \, \text{from Equation \eqref{eq:def_map_body} and }\, \mathrm{Div}\bm{P} \, \text{from Equation \eqref{eq:div_total_stress} gives} \nonumber  \\
& = -\int\limits_{ S_\mathrm{m}}\int\limits_{\eta}\left[ \bm{A}_0\cdot \delta\bm{r}+\eta \delta\lambda \bm{A}_0\cdot \bm{n}+\eta \lambda \bm{A}_0\cdot \delta\bm{n}+\eta \bm{A}_1\cdot \delta\bm{r} +\mathcal{O}(\eta^2)\right] M d\eta dS_\mathrm{m}, \\
& \quad \text{substituting} \, M \, \text{from Equation \eqref{eq:shift_ref} gives}\nonumber \\
& = -\int\limits_{ S_\mathrm{m}}\int\limits_{\eta}\left[\bm{A}_0\cdot \delta\bm{r}+\eta \delta\lambda \bm{A}_0\cdot \bm{n}+\eta \lambda \bm{A}_0\cdot \delta\bm{n}+\eta \bm{A}_1\cdot \delta\bm{r} -2\eta H\bm{A}_0\cdot \delta\bm{r}+\mathcal{O}\left(\eta^2\right)\right] d\eta dS_\mathrm{m}.
\label{eq: div P with Long H}
\end{align}
As  $\eta \in [-T/{2}, T/{2}]$ and we note that across the thickness of the shell 
\begin{eqnarray}
{\int\limits_{\eta} d\eta}=T \quad \text{and} \quad {\int\limits_{\eta} \eta d\eta}
=0.
\label{eq:homogenised_terms}
\end{eqnarray}
This allows one to simplify Equation \eqref{eq: div P with Long H} as
\begin{equation}
 -\int\limits_{{\mathcal{B}}_0}\mathrm{Div}\bm{P}\cdot\delta\bm{\chi}dV
  =\int\limits_{S_\mathrm{m}}\left[-T \big[  \widehat{\text{Div}} {\bm{P}_0}  +  \bsym{T}_1 \big] \cdot \delta\bm{r}+ \mathcal{O}\left(T^3\right)\right]dS_\mathrm{m},
  \label{eq:tot_stress_shell_var}
\end{equation}
where the mechanical traction is given by $\bsym{T}=\bm{P} \bm{N}=\bsym{T}_0 + \eta \bsym{T}_1$ and the surface divergence is defined as $\widehat{\text{Div}}  \bm{W}  = \bm{W}_{, \beta} \cdot \bm{A}^\beta $.

\bigskip
\noindent {\large \textbf{Integrals over the top and bottom boundaries in $I$}:}

The boundary terms in Equation \eqref{eq:var_inside_body} are now addressed.
The second term in Equation \eqref{eq:var_inside_body} over the top surface ($\eta = T/2, \bm{N}_\mathrm{t} = \bm{N}$) can be written using Equation \eqref{eq:total_stress_taylor}  as
\begin{align}
    \int\limits_{ S_\mathrm{t}}\bm{P} {\bm{N}}_\mathrm{t}\cdot\delta\bm{\chi}dS_\mathrm{t}
    &=
\int\limits_{ S_\mathrm{t}}\bm{P}\Big|_{\eta=T/2} {\bm{N}}_\mathrm{t}\cdot\delta\bm{\chi} \Big|_{\eta=T/2}dS_\mathrm{t}, \\
&= \int\limits_{ S_\mathrm{t}}\left[ \bsym{T}_0 \cdot\delta \bm{r}+\frac{T}{2}\delta \lambda \bsym{T}_0 \cdot \bm{n}+ \frac{T}{2} \lambda \bsym{T}_0 \cdot \delta \bm{n}+\frac{T}{2} \bsym{T}_1 \cdot\delta \bm{r}
+\mathcal{O}\left(T^2\right)\right]dS_\mathrm{t}.
\label{eq: Pnt top}
\end{align}
The above equation is evaluated over the top surface of the shell and should be transformed such that the domain of integration changes to the mid-surface.
This is achieved using the shifter relation in Equation \eqref{eq:shift_ref}, giving
\begin{equation}
    \int\limits_{ S_\mathrm{t}} \bsym{T}_0 \cdot\delta \bm{r} \, dS_t
    = \int\limits_{ S_\mathrm{m}} \bsym{T}_0 \cdot\delta \bm{r} M dS_\mathrm{m}
    = \int\limits_{ S_\mathrm{m}} \big[ \bsym{T}_0 \cdot\delta \bm{r} - TH  \bsym{T}_0 \cdot\delta \bm{r} + \mathcal{O}\left( T^2 \right) \big] dS_\mathrm{m},
\end{equation}
and applying the same procedure to the second, third, and fourth terms in Equation \eqref{eq: Pnt top} retaining only terms of $\mathcal{O}(T)$ gives
\begin{equation}
    \int\limits_{ S_\mathrm{t}} (\bullet) dS_\mathrm{t} = 
    \int\limits_{ S_\mathrm{m}} (\bullet) M dS_\mathrm{m} =
    \int\limits_{ S_\mathrm{m}} \big[  (\bullet) + \mathcal{O}\left(T^2\right)\big] dS_\mathrm{m} .
\end{equation}
Now that  the integrals have been transformed to the mid-surface, it is necessary to convert the variations of all the quantities $(\delta \lambda, \delta \bm{n})$ to $\delta \bm{r}$ in order to derive the Euler--Lagrange equations.
This is achieved by transferring the integrals to the parametric domain, applying the form of Green's theorem stated in Equation \eqref{eq:dom_to_bound} and then transforming back to the shell domain.
The second term of Equation \eqref{eq: Pnt top} with $\delta \lambda$  over $S_\mathrm{m}$ is transformed to the parametric domain and written as
\begin{align}
    \int\limits_{ S_\mathrm{m}} T & \delta \lambda \bsym{T}_0 \cdot \bm{n}dS_\mathrm{m}
    = \int\limits_{ P}T\delta \lambda \bsym{T}_0 \cdot \bm{n}{A}^{1/2} dP , \\ 
    & \quad \text{substituting} \, \delta \lambda \, \text{using Equation \eqref{eq:var_lambda}} \\
    & = -\int\limits_{ P} T \lambda  \left[\bsym{T}_0 \cdot \bm{n}\right] \bm{a}^{\alpha}\cdot\delta {\bm{a}}_{\alpha}{A}^{1/2} dP, \\
    & \quad \text{substituting} \, \delta \bm{a}_{\alpha} \, \text{from Equation \eqref{eq: delta a lpha}} \nonumber \\
    & = -\int\limits_{ P}\left[{C}^{\alpha}{A}^{1/2}\right]_{,\alpha}dP
+\int\limits_{ S_\mathrm{m}}\left[T \lambda  \left[\bsym{T}_0 \cdot \bm{n}\right] \bm{a}^{\alpha}\right]_{, \alpha}\cdot\delta \bm{r} dS_\mathrm{m}  +
\int\limits_{ P} T \lambda  \left[\bsym{T}_0 \cdot \bm{n}\right]\bm{a}^{\alpha}\left[A^{1/2}\right]_{,\alpha}\cdot\delta \bm{r} dP,  \\
& \quad \text{transforming back from parametric domain to shell domain} \nonumber \\
&= -\int\limits_{ P}\left[{C}^{\alpha}{{A}}^{1/2}\right]_{,\alpha}dP
+\int\limits_{ S_\mathrm{m}}\left[ T \lambda  \left[\bsym{T}_0 \cdot \bm{n}\right] \bm{a}^{\alpha}\right]_{, \alpha}\cdot\delta \bm{r} dS_\mathrm{m} +
\int\limits_{S_\mathrm{m}} T \lambda  \left[\bsym{T}_0 \cdot \bm{n}\right] \bm{a}^{\alpha} {\Gamma}^\beta_{\beta \alpha}\cdot\delta \bm{r} dS_\mathrm{m}, \\
&= -\int\limits_{ P}\left[{C}^{\alpha}{{A}}^{1/2}\right]_{,\alpha}dP
+\int\limits_{ S_\mathrm{m}} \widehat{\text{Div}} \left( T \lambda  \left[\bsym{T}_0 \cdot \bm{n}\right] \bm{a}^{\alpha} \otimes \bm{A}_\alpha \right) \cdot\delta \bm{r} dS_\mathrm{m} .
\end{align}
%
The first term containing the derivative over the parametric domain is transformed using the modified Green's theorem
\eqref{eq:dom_to_bound} as 
\begin{equation}
    -\int\limits_{ P}\left[{C}^{\alpha}{A}^{1/2}\right]_{,\alpha}dP = -\int\limits_{C_\mathrm{m}}{C}^{\alpha} {\nu}_\alpha dl
=-\int\limits_{C_\mathrm{m}}T \lambda  \left[\bsym{T}_0 \cdot \bm{n}\right] \overline{\bm{\nu}} \cdot\delta \bm{r}dl
=-\int\limits_{C_\mathrm{m} \setminus C_\mathrm{m}^\mathrm{u}}T \lambda  \left[\bsym{T}_0 \cdot \bm{n}\right] \overline{\bm{\nu}}
 \cdot\delta \bm{r} dl .
\end{equation}

The third term of Equation \eqref{eq: Pnt top} with $\delta \bm{n}$ is transformed by using Equation \eqref{eq:var_def_nor} as
\begin{align}
    \int\limits_{ S_\mathrm{m}}T \lambda \bsym{T}_0 \cdot \delta \bm{n}dS_\mathrm{m}
    &=\int\limits_{ P} T \lambda \bsym{T}_0 \cdot \delta \bm{n}{A}^{1/2} dP =-\int\limits_{P}T\lambda \big[\bsym{T}_0 \cdot\bm{a}^\alpha\big] \big[ \bm{n} \cdot \delta \bm{a}_\alpha \big] {A}^{1/2}dP =0 . 
\end{align}
The above term vanishes on account of the Kirchhoff-Love assumption of zero shear stresses, i.e., $\bsym{T}_0 \cdot\bm{a}^\alpha =0$.

%

Application of similar mathematical manipulations  to the second and third terms in Equation \eqref{eq:var_inside_body} corresponding to the top and bottom boundaries and neglecting the higher order terms results in
\begin{align}
    \int\limits_{ S_\mathrm{t}}\bm{P}{\bm{N}}_\mathrm{t}\cdot\delta\bm{\chi}dS_\mathrm{t}
    &
 +\int\limits_{ S_\mathrm{b}}\bm{P}{\bm{N}}_\mathrm{b}\cdot\delta\bm{\chi}dS_\mathrm{b} \nonumber \\
 =& \int\limits_{ {\mathcal{S}}_\mathrm{m}}\Bigg[
    \widehat{\text{Div}} \left( T \lambda  \left[\bsym{T}_0 \cdot \bm{n}\right] \bm{a}^{\alpha} \otimes \bm{A}_\alpha \right)
    + T \bsym{T}_1 
-2TH \bsym{T}_0
\Bigg]\cdot \delta \bm{r}d{\mathcal{S}}_\mathrm{m} \nonumber \\
 & - \int\limits_{{\mathcal{C}}_\mathrm{m} \setminus {\mathcal{C}}_\mathrm{m}^\mathrm{u} } T \lambda  \left[\bsym{T}_0 \cdot \bm{n}\right] \overline{\bm{\nu}} 
\cdot \delta \bm{r}dl .
\label{eq: top bottom surf}
\end{align}

{\large \textbf{Integral over the lateral boundary in $I$}:}

The fourth term in Equation \eqref{eq:var_inside_body} is the contribution of Piola stress over the lateral boundary.
This integral can be expressed using Equations \eqref{eq:homogenised_terms}, \eqref{eq:lateral_to_mid}, and \eqref{eq:lateral_nor_to_mid}
as
%
%
\begin{equation}
\int\limits_{ S_\mathrm{l}}\bm{P} \bm{\nu}_\text{l}\cdot\delta\bm{\chi}dS_\mathrm{l}
=\int\limits_{C_\mathrm{m}}\left[T\bm{P}_0 \bm{\nu} \cdot \delta \bm{r}+\mathcal{O}\left(T^3\right)\right]dl 
=\int\limits_{C_\mathrm{m} \setminus C_\mathrm{m}^\mathrm{u} }\left[T \bm{P}_0 \bsym{\nu}  \cdot \delta \bm{r}+\mathcal{O}\left(T^3\right)\right]dl.
\label{eq: lateral boundary integ result}
\end{equation}

{\large \textbf{Total contribution}:}

Upon combining Equations \eqref{eq:tot_stress_shell_var}, \eqref{eq: top bottom surf} and \eqref{eq: lateral boundary integ result}, and neglecting the higher order terms in $T$, 
the entire contribution from the Piola-stress from Equation \eqref{eq:var_inside_body}  is given by 
\begin{empheq}[box=\fbox]{align}
I
 =  \int\limits_{ {\mathcal{S}}_\mathrm{m}}\Bigg[-T \, \widehat{\text{Div}}  \bm{P}_0  + \widehat{\text{Div}} \left(  T \lambda  \left[\bsym{T}_0 \cdot \bm{n}\right] \bm{a}^{\alpha} \otimes \bm{A}_\alpha \right) 
    -
2TH\bsym{T}_0 \Bigg]\cdot \delta \bm{r}d{\mathcal{S}}_\mathrm{m} 
+\int\limits_{{\mathcal{C}}_\mathrm{m} \setminus {\mathcal{C}}_\mathrm{m}^\mathrm{u} } T \Big[{\bm{P}_0} \bsym{\nu} - \lambda  \left[\bsym{T}_0 \cdot \bm{n}\right] \bm{\nu
} \Big]  \cdot \delta \bm{r}dl 
\label{eq: final contribution piola stress}
\end{empheq}



\subsubsection{Terms involving the external loads}
\label{sec: external loads terms}

The terms involving the external pressure, traction, and body force in Equation \eqref{eq:var_2} involve the variation $\delta \bm{\chi}$.
In order to derive the shell equations, they need to be expressed in terms of the variation $\delta \bm{r}$.
It is noted that one of the main novelties in the present work is the differentiation between the pressure load applied at the top and bottom surfaces.
Since the pressure is always applied in the current configuration, the integrals need to be transformed to the reference and the parametric domain for application of the Green's theorem.

Consider the second to last term in Equation \eqref{eq:var_2} involving pressure on the top surface ($\eta = T/2, \bm{n}_t = \bm{n}$). It is written using  Equations \eqref{eq:press_term_forms}, \eqref{eq:ref_top_bot_to_mid}, and  \eqref{eq:def_surface_top_bottom} as
\begin{align}
    \int\limits_{ s_\mathrm{t}}\bm{p}_\mathrm{t}\cdot \delta\bm{\chi} ds_\mathrm{t}
    =&
    -\int\limits_{ S_\mathrm{t}}p_\mathrm{t}\bm{n} \cdot\delta\bm{\chi} \Big|_{\eta=T/2}
\mu\Big|_{\eta=T/2}
    \frac{{\widehat{a}}^{1/2}} {M}
   dS_\mathrm{t} \\
   & \quad \text{using Equations \eqref{eq:shift_ref} and \eqref{eq:var_chi_1}}  \nonumber \\
   =& -\int\limits_{ S_\mathrm{t}}p_\mathrm{t}\bm{n}\cdot\left[\delta \bm{r}+\delta\lambda \frac{T}{2}\bm{n}+ \frac{T}{2}  \lambda  \delta\bm{n}\right]
\left[1-T\left[\lambda h + H \right]+\mathcal{O}\left(T^2\right)\right]
{\widehat{a}}^{1/2} dS_\mathrm{t}, \\
& \quad \text{using} \; \bm{n} \cdot \bm{n}=1, \; \bm{n} \cdot \delta \bm{n} = 0 , \; \lambda \widehat{a}^{1/2} = 1 \nonumber \\
= & \int\limits_{ S_\mathrm{t}}\left[-{\lambda}^{-1}p_\mathrm{t}\bm{n}\cdot\delta \bm{r}+\bm{a}^{\alpha}\cdot \delta \bm{a}_{\alpha} \frac{T}{2}p_\mathrm{t}+T \left[ h + \lambda^{-1} H\right] p_\mathrm{t}\bm{n}\cdot\delta \bm{r}+\mathcal{O}\left(T^2\right)\right] dS_\mathrm{t}.
\end{align}

Upon defining the average pressure $\overline{p}=[p_\mathrm{t}+p_\mathrm{b}]/2$, the total contribution of pressure to the top and bottom surfaces in Equation \eqref{eq:var_2} can be written as
\begin{eqnarray}
\int\limits_{ s_\mathrm{t}}\bm{p}_\mathrm{t}\cdot \delta\bm{\chi}ds_\mathrm{t}+\int\limits_{ s_\mathrm{b}}\bm{p}_\mathrm{b}\cdot \delta\bm{\chi}ds_\mathrm{b}&=&
 \int\limits_{ \mathcal{S}_\mathrm{t}}\left[-{\lambda}^{-1}p_t\bm{n}\cdot\delta \bm{r}
 +T \left[ h + \lambda^{-1}H\right] p_t\bm{n}\cdot\delta \bm{r}+\mathcal{O}\left(T^2\right)\right] dS_\mathrm{t}
\nonumber\\
&+&\int\limits_{ S_\mathrm{b}}\left[{\lambda}^{-1}p_\mathrm{b}\bm{n}\cdot \delta \bm{r}
+T\left[ h + \lambda^{-1} H\right]p_\mathrm{b}\bm{n}\cdot \delta \bm{r}+\mathcal{O}\left(T^2\right)\right]dS_\mathrm{b}
\nonumber\\
&+&\int\limits_{ S_\mathrm{m}}\left[
\bm{a}^{\alpha}\cdot \delta \bm{a}_{\alpha} T\overline{p}
+\mathcal{O}\left(T^2\right)\right]dS_\mathrm{m} .
\label{eq:cont_press_1}
\end{eqnarray}
The only term above that needs further modification is the one with $\delta \bm{a}_\alpha$.
This is achieved by pulling it back to the parametric domain and applying the Green's theorem to get
%
\begin{eqnarray}
\int\limits_{ S_\mathrm{m}}\bm{a}_{\alpha}\cdot \delta \bm{a}^{\alpha} T\overline{p} dS_\mathrm{m}&=&\int\limits_{ P} T \overline{p} \bm{a}^{\alpha} \cdot \delta \bm{a}_{\alpha}  {A}^{1/2}dP, \nonumber \\
&=&\int\limits_{ P}\left[{{E}}^{\alpha}A^{1/2}\right]_{,\alpha}dP
-\int\limits_{ S_\mathrm{m}} \Big[ \left[T\overline{p} \bm{a}^{\alpha} \right]_{, \alpha} + T\overline{p} \bm{a}^{\alpha} {\Gamma}^\beta_{\beta \alpha} \Big] \cdot\delta \bm{r}dS_\mathrm{m}, \nonumber \\
&=& \int\limits_{ P}\left[{{E}}^{\alpha}A^{1/2}\right]_{,\alpha}dP
- \int\limits_{ S_\mathrm{m}} \widehat{\text{Div}} \left( T\overline{p} \bm{a}^{\alpha} \otimes \bm{A}_\alpha \right) \cdot \delta \bm{r} dS_\mathrm{m}
\label{eq: a alpha term modified}
\end{eqnarray}
with ${E}^{\alpha}=   T\overline{p} \bm{a}^{\alpha} \cdot \delta \bm{r} $.
An application of the modified Green's theorem \eqref{eq:dom_to_bound} to the first term  results in
\begin{equation}
\int\limits_{ P}\left[{{E}}^{\alpha} A^{1/2}\right]_{,\alpha}dP
=\int\limits_{C_\mathrm{m}}{E}^{\alpha} {\nu}_{\alpha}dl
=\int\limits_{C_\mathrm{m}} T\overline{p} \overline{\bm{\nu}} \cdot \delta \bm{r}dl
=\int\limits_{C_\mathrm{m}\setminus C_\mathrm{m}^\mathrm{u}} T\overline{p} \overline{\bm{\nu}} \cdot \delta \bm{r}dl.
\label{eq: converting Ealpha term}
\end{equation}

Upon substituting the Equations \eqref{eq: a alpha term modified} and \eqref{eq: converting Ealpha term} back into Equation \eqref{eq:cont_press_1}, using the shifter relations \eqref{eq:ref_top_bot_to_mid} to transform all integrals to the mid-surface, and neglecting the $\mathcal{O}(T^2)$ terms, the total contribution due to the pressure loads can be written as
\begin{align}
\int\limits_{ s_\mathrm{t}}\bm{p}_\mathrm{t}\cdot \delta\bm{\chi}ds_\mathrm{t}+ \int\limits_{ s_\mathrm{b}}\bm{p}_\mathrm{b}\cdot \delta\bm{\chi}ds_\mathrm{b}
=\int\limits_{ {\mathcal{S}}_\mathrm{m}}\Big[-{\widehat{a}}^{1/2}\widetilde{p}\bm{n}\cdot \delta \bm{r}+2T H \overline{p}\bm{n}\cdot \delta \bm{r}- \widehat{\text{Div}} \left( T\overline{p} \bm{a}^{\alpha} \otimes \bm{A}_\alpha \right)
 \cdot \delta \bm{r} \Big]d{\mathcal{S}}_\mathrm{m} \nonumber \\
 +\int\limits_{{\mathcal{C}}_\mathrm{m}\setminus {\mathcal{C}}_\mathrm{m}^\mathrm{u}}T\overline{p} \overline{\bm{\nu}} \cdot \delta \bm{r}dl.
 \label{eq:total_press_term}
\end{align}
where  the difference in the top and bottom pressure is defined by $ \widetilde{p}=p_\mathrm{t}-p_\mathrm{b}$.

The contribution due to the dead load traction on the Neumann boundary of the shell's mid-surface is written as
\begin{eqnarray}
\int\limits_{ C_\mathrm{m} \setminus C_\mathrm{m}^\mathrm{u}}\bm{t}_\ell \cdot\delta\bm{\chi}dl =\int\limits_{ C_\mathrm{m} \setminus C_\mathrm{m}^\mathrm{u}}\bm{t}_\ell \cdot\delta\bm{\chi} \Big|_{\eta=0}dl
=\int\limits_{ C_\mathrm{m} \setminus C_\mathrm{m}^\mathrm{u}}\bm{t}_\ell \cdot\delta \bm{r}dl.
\label{eq:dead_load_trac_cont}
\end{eqnarray}
The virtual work due to the body force per unit volume can be further simplified as
\begin{eqnarray}
\int\limits_{{\mathcal{B}}_0}\bm{\mathfrak{B}}\cdot \delta\bm{\chi} dV
&=&\int\limits_{ S_\mathrm{m}}\left[T\bm{\mathfrak{B}}_0\cdot \delta \bm{r}+\mathcal{O}\left(T^3\right)\right]dS_\mathrm{m},
\label{eq:body_force_cont}
\end{eqnarray}
with $\bm{\mathfrak{B}} (\eta, \theta^\alpha)=\bm{\mathfrak{B}}_0 (\theta^\alpha) +\eta \bm{\mathfrak{B}}_1 (\theta^\alpha) +\bm{\mathcal{O}}(\eta^2)$ in ${\mathcal{B}}_0$.

\bigskip
{\large \textbf{Total contribution}:}

Considering  Equations \eqref{eq:total_press_term}, \eqref{eq:dead_load_trac_cont}, and \eqref{eq:body_force_cont}, 
and neglecting the higher order terms,
the total contribution due to the external loads can be written in terms of the variation $\delta \bm{r}$ as
%
%
\begin{empheq}[box=\fbox]{align}
- & \int\limits_{ s_\mathrm{t}}\bm{p}_\mathrm{t}\cdot \delta\bm{\chi}ds_\mathrm{t} -\int\limits_{ s_\mathrm{b}}\bm{p}_\mathrm{b}\cdot \delta\bm{\chi}ds_\mathrm{b}- \int\limits_{{\mathcal{B}}_0}\bm{\mathfrak{B}}\cdot \delta\bm{\chi} dV-\int\limits_{C_\mathrm{m} \setminus C_\mathrm{m}^\mathrm{u}}\bm{t}_\ell \cdot\delta\bm{\chi}dl
\nonumber \\
=&\int\limits_{ {\mathcal{S}}_\mathrm{m}}\left[
\big[  {\widehat{a}}^{1/2}\widetilde{p} -2T H \overline{p} \big] \bm{n}
+\widehat{\text{Div}} \left( T\overline{p} \bm{a}^{\alpha} \otimes \bm{A}_\alpha \right)
+ T\bm{\mathfrak{B}}_0 \right] \cdot \delta \bm{r} d{\mathcal{S}}_\mathrm{m}
-\int\limits_{C_\mathrm{m}\setminus C_\mathrm{m}^\mathrm{u}}\left[T\overline{p}  \overline{\bm{\nu}} +\bm{t}_\ell \right]\cdot \delta \bm{r}dl  .
\label{eq:total_cont_ext}
\end{empheq}

\subsubsection{Shell equations in weak form}
Starting with the first variation of the total energy functional in Equation \eqref{eq:var_2},  the total contribution due to the Piola stress terms (internal virtual work) were derived in Section \ref{sec: piola terms} and the total contribution due to the pressure, traction, and body force in Section \ref{sec: external loads terms}.
Combining the total contributions in Equations \eqref{eq: final contribution piola stress} and \eqref{eq:total_cont_ext},  the  first variation of energy can be written in terms of the shell variables as
\begin{empheq}[box=\fbox]{align}
   \delta \widetilde{\Pi}[\bm{r},p_0; \delta \bm{r}, \delta p_0]
    =& \int\limits_{ S_\mathrm{m}}\Bigg[-T \, \widehat{\text{Div}}  \bm{P}_0  +\widehat{\text{Div}} \left(  T \widetilde{P} \bm{a}^{\alpha}   \otimes \bm{A}_\alpha \right)  -
2TH\bsym{T}_0
\nonumber \\
& + \big[  {\widehat{a}}^{1/2}\widetilde{p} -2T H \overline{p} \big] \bm{n}
+ T\bm{\mathfrak{B}}_0 
\Bigg]\cdot \delta \bm{r}d{\mathcal{S}}_\mathrm{m} -\int\limits_{ S_\mathrm{m}}T\left[J_0-1\right]\delta p_0 dS_\mathrm{m}\nonumber \\ 
&+\int\limits_{{\mathcal{C}}_\mathrm{m} \setminus {\mathcal{C}}_\mathrm{m}^\mathrm{u} } 
\bigg[ T \Big[ \bsym{P}_0 {\bsym{\nu}} - \widetilde{P} \overline{\bm{\nu}} \Big] -\bm{t}_\ell \bigg] 
\cdot \delta \bm{r}dl . 
 \label{eq:FE_lin_dis}
\end{empheq}
Here $\widetilde{P} = \overline{p} + \lambda [\bsym{T}_0 \cdot \bsym{n}]$ can be physically interpreted as the ``squeezing'' pressure on the shell wall.
The necessary governing equations and boundary conditions are obtained by setting $\delta \widetilde{\Pi} = 0$.

The formulation is expressed in terms of two unknowns, the displacement vector $\bm{r}$ of the shell mid-surface  and the zeroth order term of the Lagrange multiplier $p_0$ on the shell mid-surface.
Since the top and bottom surfaces were treated separately to make the model ``embedding-aware'' by accounting for the appropriate boundary conditions, several non-standard terms arise in comparison to the classical thin shell models \citep{Cirak:2001aa, kiendl2015isogeometric, liu2022computational}.
In particular, the pressure on the top and bottom surface can be distinguished due to the terms $\widetilde{p}$ and $\overline{p}$ as opposed to a single variable for pressure at the mid-surface as appears in classical models.
The effect of Piola stress is directly accounted for by the terms involving $\bm{P}_0$ and its derivatives as opposed to taking a stress resultant by integrating $\bm{P}$ through the thickness in classical models.
The influence of these terms in this novel model are discussed with the aid of examples in  Section \ref{sec: numerical}.

\subsection{Shell governing equations and boundary conditions}
\label{sec: strong form}
The necessary partial differential equations for the shell model are obtained as the Euler--Lagrange equations by vanishing of the first variation $\delta \widetilde{\Pi}= 0$ in Equation \eqref{eq:FE_lin_dis}.
Since the variations $\delta \bm{r}$ and $\delta p_0$ are arbitrary, the following PDE is obtained over the  mid-surface of the shell:
\begin{align}
    T  \, \widehat{\text{Div}}  \bm{P}_0  & - \widehat{\text{Div}} \left(  T \Big[  \lambda  \left[\bsym{T}_0 \cdot \bm{n}\right] \bm{a}^{\alpha} +  \overline{p} \bm{a}^{\alpha}  \Big] \otimes \bm{A}_\alpha \right) \nonumber \\
&+ 
2TH\bsym{T}_0
- \big[  {\widehat{a}}^{1/2}\widetilde{p} -2T H \overline{p} \big] \bm{n}
+ T\bm{\mathfrak{B}}_0 =\bm{0} \quad \forall \bm{X} \in S_\mathrm{m},
\label{eq: shell gov eqn}
\end{align}
along with the incompressibility condition
\begin{equation}
    J_0 - 1 = 0 \quad \forall \bm{X} \in S_\mathrm{m},
    \label{eq:incomp_surf}
\end{equation}
and the boundary condition on the Neumann boundary of the mid-surface
\begin{equation}
T \Big[ \bsym{P}_0 {\bsym{\nu}} - \widetilde{P} \overline{\bm{\nu}}  \Big] -\bm{t}_\ell
=\bm{0} \quad \forall \bm{X} \in {\mathcal{C}}_\mathrm{m} \setminus {\mathcal{C}}_\mathrm{m}^\mathrm{u} .
\label{eq: strong bc}
\end{equation}

The governing Equation \eqref{eq: shell gov eqn} can be simplified by assuming uniform thickness in the reference configuration and employing the plane-stress approximation $\bsym{T}_0\cdot \bm{n}=0$. 
Application of the plane-stress condition yields an explicit expression for the Lagrange multiplier $p_0$ in terms of the deformation gradient and the chosen hyperelastic energy density function.
By recognising that the Piola stress can be decomposed as $\bm{P}_0={{P}_0}^{\alpha \beta}  \bm{a}_\alpha \otimes \bm{A}_\beta + {P_0}_\perp \bm{n} \otimes \bm{N}$ where ${P_0}_\perp=0$ due to the plane-stress approximation, the equations for in-plane and normal directions can be separated and when written in  index notation take the form
\begin{empheq}[box=\fbox]{align}
     {{P}_0}^{\gamma \delta}_{, \delta} + {{P}_0}^{\gamma \delta} {\Gamma}^\alpha_{\delta \alpha}   -    {{P}_0}^{\zeta \delta} \bm{a}_\zeta \cdot \bm{a}^\gamma_{, \delta} -      \bigg[ \overline{p}_{, \alpha} +   \Big[ {\Gamma}^\beta_{\beta \alpha} - {\gamma}^\beta_{\beta \alpha} \Big] \overline{p} \bigg]  a^{\alpha \gamma} &= 0 \quad \forall \bm{X} \in S_\mathrm{m}  , \label{eq:strong1} \\
    {{P}_0}^{\gamma \delta} a_{\gamma \beta} b^{\ \beta}_\delta + \widehat{a}^{1/2} T^{-1} \widetilde{p} + 2 [h-H] \overline{p}   &= 0 \quad \forall \bm{X} \in S_\mathrm{m}.
    \label{eq:strong2}
\end{empheq}

Here we note the departure from  classical thin shell formulations where the PDEs are presented in a homogenised form by integrating over the thickness and defining average variables \citep{kiendl2015isogeometric}.
There equations are expressed  in terms of bending moment and normal forces, with their conjugates representing the change in curvature of the deforming body and the virtual strain.

The equations derived here are used in the subsequent section to study the response of nonlinear thin shells.
The systems under consideration reflect the cases of a thin plate, without any reference curvature of the mid-surface, and  an infinite cylinder with one nonzero component of the curvature tensor when expressed in the local triads.

\section{Illustrative examples}
\label{sec: numerical}
The primary aim of this section is to illustrate the theory of hyperelastic thin shells derived in Section \ref{sec: shell eqns from 3D}.
The Euler--Lagrange equations presented in Section \ref{sec: strong form} are used to derive the response equations and solve the boundary value problems for the deformation of a flat plate and a cylindrical shell.
For analytical derivations, a simple form of an  incompressible neo-Hookean strain energy denstity function is used:
\begin{equation}
    W = \mu [I_1 - 3],
\end{equation}
where $\mu$ is the shear modulus.
For this model, the Piola stress is calculated as
%
\begin{equation}
    \bm{P}= \mu \bm{F} -p\bm{F}^{-\mathrm{T}},
\end{equation}
and its zeroth order component from Equation \eqref{eq:total_stress_taylor} is written in terms of the local triad of the shell mid-surface as
\begin{eqnarray}
    \bm{P}_0= \left[\mu A^{\alpha \beta } -p_0 a^{\alpha \beta }\right] \bm{a}_\alpha \otimes \bm{A}_\beta + \left[\mu \lambda - \frac{p_0}{\lambda}\right] \bm{n} \otimes \bm{N}.
    \label{eq: examples neo hooke stress}
\end{eqnarray}
The plane stress condition $\bsym{T}_0 \cdot \bm{n}=0 $ leads to the explicit determination of the Lagrange multiplier as
\begin{eqnarray}
   p_0= \mu \lambda^2.
   \label{eq: p_0 final value}
\end{eqnarray}

\subsection{Thin plate under boundary traction and pressure across surfaces}
\begin{figure}
  \begin{center}
  \subfigure[]{ \def\svgwidth{8cm} 
\begingroup%
  \makeatletter%
  \providecommand\color[2][]{%
    \errmessage{(Inkscape) Color is used for the text in Inkscape, but the package 'color.sty' is not loaded}%
    \renewcommand\color[2][]{}%
  }%
  \providecommand\transparent[1]{%
    \errmessage{(Inkscape) Transparency is used (non-zero) for the text in Inkscape, but the package 'transparent.sty' is not loaded}%
    \renewcommand\transparent[1]{}%
  }%
  \providecommand\rotatebox[2]{#2}%
  \newcommand*\fsize{\dimexpr\f@size pt\relax}%
  \newcommand*\lineheight[1]{\fontsize{\fsize}{#1\fsize}\selectfont}%
  \ifx\svgwidth\undefined%
    \setlength{\unitlength}{292.74209578bp}%
    \ifx\svgscale\undefined%
      \relax%
    \else%
      \setlength{\unitlength}{\unitlength * \real{\svgscale}}%
    \fi%
  \else%
    \setlength{\unitlength}{\svgwidth}%
  \fi%
  \global\let\svgwidth\undefined%
  \global\let\svgscale\undefined%
  \makeatother%
  \begin{picture}(1,0.68260317)%
    \lineheight{1}%
    \setlength\tabcolsep{0pt}%
    \put(0,0){\includegraphics[width=\unitlength,page=1]{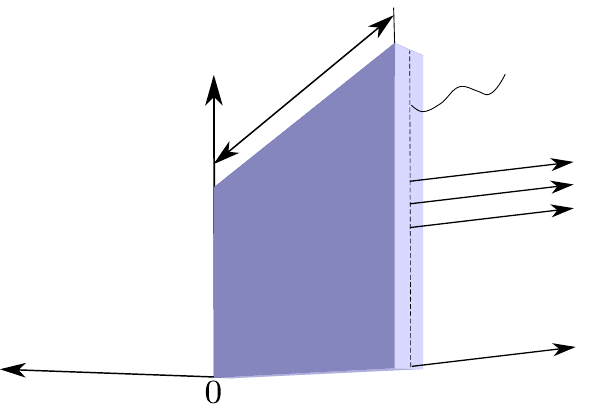}}%
    \put(0.48889006,0.64163809){\color[rgb]{0,0,0}\makebox(0,0)[lt]{\lineheight{1.25}\smash{\begin{tabular}[t]{l}$L$\end{tabular}}}}%
    \put(0.75487762,0.58074476){\color[rgb]{0,0,0}\makebox(0,0)[lt]{\lineheight{1.25}\smash{\begin{tabular}[t]{l}$\mathrm{midsurface}$\end{tabular}}}}%
    \put(0.83249759,0.26831833){\color[rgb]{0,0,0}\makebox(0,0)[lt]{\lineheight{1.25}\smash{\begin{tabular}[t]{l}$t_\ell$\end{tabular}}}}%
    \put(0.17399623,0.56723154){\color[rgb]{0,0,0}\makebox(0,0)[lt]{\lineheight{1.25}\smash{\begin{tabular}[t]{l}$x_2, \ X_2$\end{tabular}}}}%
    \put(0.04244555,0.00987516){\color[rgb]{0,0,0}\makebox(0,0)[lt]{\lineheight{1.25}\smash{\begin{tabular}[t]{l}$x_3, \ X_3$\end{tabular}}}}%
    \put(0.74678067,0.00987519){\color[rgb]{0,0,0}\makebox(0,0)[lt]{\lineheight{1.25}\smash{\begin{tabular}[t]{l}$x_1, \ X_1$\end{tabular}}}}%
  \end{picture}%
\endgroup%
}
  \subfigure[]{ \def\svgwidth{8cm} 
\begingroup%
  \makeatletter%
  \providecommand\color[2][]{%
    \errmessage{(Inkscape) Color is used for the text in Inkscape, but the package 'color.sty' is not loaded}%
    \renewcommand\color[2][]{}%
  }%
  \providecommand\transparent[1]{%
    \errmessage{(Inkscape) Transparency is used (non-zero) for the text in Inkscape, but the package 'transparent.sty' is not loaded}%
    \renewcommand\transparent[1]{}%
  }%
  \providecommand\rotatebox[2]{#2}%
  \newcommand*\fsize{\dimexpr\f@size pt\relax}%
  \newcommand*\lineheight[1]{\fontsize{\fsize}{#1\fsize}\selectfont}%
  \ifx\svgwidth\undefined%
    \setlength{\unitlength}{294.0845693bp}%
    \ifx\svgscale\undefined%
      \relax%
    \else%
      \setlength{\unitlength}{\unitlength * \real{\svgscale}}%
    \fi%
  \else%
    \setlength{\unitlength}{\svgwidth}%
  \fi%
  \global\let\svgwidth\undefined%
  \global\let\svgscale\undefined%
  \makeatother%
  \begin{picture}(1,0.62516649)%
    \lineheight{1}%
    \setlength\tabcolsep{0pt}%
    \put(0,0){\includegraphics[width=\unitlength,page=1]{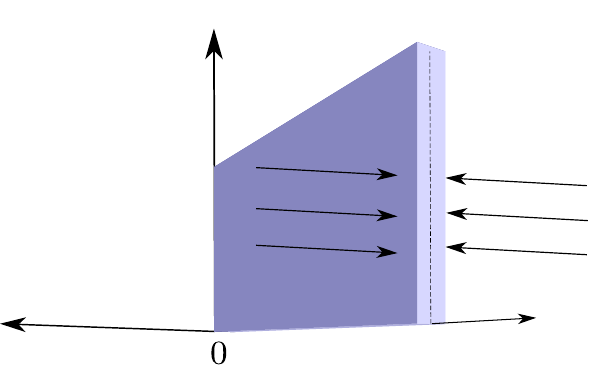}}%
    \put(0.01382463,0.00983009){\color[rgb]{0,0,0}\makebox(0,0)[lt]{\lineheight{1.25}\smash{\begin{tabular}[t]{l}$x_3, \ X_3$\end{tabular}}}}%
    \put(0.77176299,0.00983011){\color[rgb]{0,0,0}\makebox(0,0)[lt]{\lineheight{1.25}\smash{\begin{tabular}[t]{l}$x_1, \ X_1$\end{tabular}}}}%
    \put(0.20808676,0.59473738){\color[rgb]{0,0,0}\makebox(0,0)[lt]{\lineheight{1.25}\smash{\begin{tabular}[t]{l}$x_2, \ X_2$\end{tabular}}}}%
    \put(0.82731554,0.35043507){\color[rgb]{0,0,0}\makebox(0,0)[lt]{\lineheight{1.25}\smash{\begin{tabular}[t]{l}$p$\end{tabular}}}}%
    \put(0.44432083,0.36531184){\color[rgb]{0,0,0}\makebox(0,0)[lt]{\lineheight{1.25}\smash{\begin{tabular}[t]{l}$p$\end{tabular}}}}%
  \end{picture}%
\endgroup%
}
  \subfigure[]{
  \def\svgwidth{8.5cm} 
 \input{p11.tex}
 }
 \subfigure[]{\def\svgwidth{8.77cm} 
 \input{p77.tex}}
   \caption{
   (a) Schematic of a thin plate with roller support at $X_1=0$ and a uniform traction applied at $X_1 = L$.
   (b) Schematic of a thin plate with equal and opposite pressure $p$ applied on the two surfaces at $X_3 = \pm T/2$.
   (c) Response of the plate under traction governed by Equation \eqref{eq: traction response eq} for two thickness values ($\Grave{T} = \{ 1/20, 1/30\}$).
   (d) Response of the plate squeezed by equal and opposite pressure governed by Equation \eqref{eq: pressure response eq}.
   }
      \label{fig:res_1}
  \end{center}
\end{figure}

Consider the problem of the deformation of a thin hyperelastic plate with sides of length $L$ and a uniform thickness $T$ in its reference configuration.
Two types of loading scenarios are analysed as shown in Figure \ref{fig:res_1}(a,b).
In the first case, a uniformly distributed traction per unit area $t_\ell$ is applied at the boundary $X_1 = L$ along the $X_1$ axis; while the points on the boundary $X_1=0$ are free to move along the $X_2$ axis but their motion along the $X_1$ direction is constrained.
In the second case, the plate is squeezed by a uniform pressure $p$ on its two surfaces at $X_3 = \pm T/2$.

The homogeneous deformation of the plate can be represented by the 3-dimensional deformation gradient $[\bm{F}] = \text{diag}(\lambda_1, \lambda_2, \lambda)$.
Considering the mid-surface of the shell, the covariant metric tensors in the reference and the deformed configurations are given by \begin{eqnarray}
    \left[{A}_{\alpha \beta}\right]=\begin{bmatrix}
1 & 0 \\
0 & 1 
\end{bmatrix} \quad  \text{and} \quad
  \left[{a}_{\alpha \beta}\right]=\begin{bmatrix}
\lambda^2_1 & 0 \\
0 & \lambda^2_2 
\end{bmatrix},
\label{eq:plate_A_a}
\end{eqnarray}
with the determinants $A=1$ and $a = \lambda_1^2 \lambda_2^2$.
The curvature tensor vanishes for the flat plate. 
The incompressibility condition \eqref{eq:incomp_surf} gives the thickness stretch as $\lambda = \sqrt{A/a}$ and imposes the constraint $\lambda \lambda_1 \lambda_2 = 1$.


For the first problem with traction, Equations \eqref{eq:strong1} and \eqref{eq:strong2} are trivially satisfied due to homogeneous deformation.
The boundary condition \eqref{eq: strong bc} applied on the traction-free boundary $X_2 = \{ 0, L\}, \bm{\nu} = \pm \bm{e}_2$ along with Equation \eqref{eq: p_0 final value} results in the relation
\begin{equation}
    \lambda_2 = \lambda.
\end{equation}
Application of  Equation \eqref{eq: strong bc} on the boundary $X_1 = L, \bm{\nu} = \bm{e}_1$
results in the response equation
\begin{eqnarray}
     \frac{t_\ell}{\mu L}=\grave{T}\left[\lambda_1-\lambda_1^{-2}\right] = \grave{T}\left[\lambda_2^{-2}-\lambda_2^{4}\right],
     \label{eq: traction response eq}
\end{eqnarray}
where the dimensionless thickness $\grave{T}=\dfrac{T}{L}$ is defined in order to incorporate the effect of thickness on the behaviour of the system.
This response is plotted in Fig. \ref{fig:res_1}(c). It is observed that as traction $t_\ell$ is applied on one boundary while the other boundary remains free, the thin plate deforms exhibiting a monotonic increase in $\lambda_1$ and a monotonic decrease in $\lambda_2$.
The effect of the thickness of the thin plate is also depicted, and it is observed that for a thinner case, the system is easier to deform.



For the second problem where an equal and opposite pressure $p$ is applied on the two surfaces of the plate in the deformed configuration, symmetry and incompressibility lead to the condition
\begin{equation}
    \lambda_1 = \lambda_2 = \lambda^{-2}.
\end{equation}
The equilibrium Equations \eqref{eq:strong1} and \eqref{eq:strong2} are trivially satisfied due to homogeneity of the deformation. Substituting $\bm{t}_\ell = \bm{0}$ and $\overline{p} = p$ in the boundary condition \eqref{eq: strong bc} results in 
\begin{equation}
    \frac{p}{\mu}=\lambda_1^2-\lambda_1^{-4} = \lambda_2^2-\lambda_2^{-4}.
    \label{eq: pressure response eq}
\end{equation}
This response is plotted in Figure \ref{fig:res_1}(d), and it is evident that as the applied pressure increases, the stretch $\lambda_1$ increases monotonically.
It is noteworthy that the shell model is able to capture this effect even with the plane-stress assumption in place.
Several classical thin shell models \citep{Cirak:2001aa, kiendl2015isogeometric, liu2022computational} are unable to capture this loading condition because they consider an average pressure acting directly on the mid-surface.
It is noted that the exact same response Equations \eqref{eq: traction response eq} and \eqref{eq: pressure response eq} are obtained from the equations of 3D elasticity  \citep{ogden1997non} if the plate is considered as a 3D block, thereby validating the current model.

 

\subsection{Inflation of a thin cylindrical shell}

\begin{figure}
 \begin{center}
 \subfigure[]{
   \includegraphics[width=0.2\linewidth]{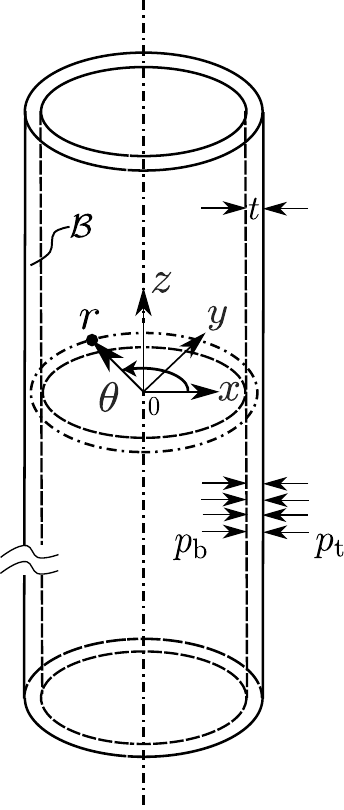}}
   \subfigure[]{ \def\svgwidth{10.82cm} 
 \input{p99.tex}}
    \caption{
(a) A schematic  showing the deformed configuration of an infinite cylindrical shell with thickness $t$ under the influence of pressures, $p_b$ and $p_t$, at the inner and outer surfaces respectively. 
(b) Variation of the mid-surface azimuthal stretch ($\lambda_\theta$) with the dimensionless pressure difference $\Delta p = [p_b-p_t]/\mu$ for the cylinder modelled as a thin shell and a 3D solid.
The shell response can be controlled by changing the pressure on the outer surface $p_t/\mu$ and the response for the 3D model lies within the region $p_t/\mu \in (0, 0.08)$.
}
      \label{fig:res_2}
  \end{center}
\end{figure}

Consider an infinitely long thin hyperelastic cylindrical shell as shown in Figure \ref{fig:res_2}(a).
Different magnitudes of pressure are applied at the internal and the external surfaces of the shell while a unit stretch is maintained along the $z$ axis resulting in an axisymmetric inflation.
The axisymmetric deformation of a cylindrical shell can be written as
\begin{equation}
    \theta = \Theta, \quad \quad 
z = Z, \quad \quad
\bm{r} = \bm{R} + \bm{u} = \left[R+u\right]\bm{e}_{\rho} + Z\bm{e}_z = r\bm{e}_{\rho} + Z\bm{e}_z
\end{equation}
Here, $\theta$ and $z$ denote the deformed coordinates corresponding to their azimuthal and axial counterparts in the reference configuration ($\Theta$ and $Z$, respectively).
The unit vectors along the axial and radial directions are represented by $\bm{e}_z$ and $\bm{e}_{\rho}$, respectively.
Moreover, $R$ and $r$ identify the radius at the mid-surface of the cylindrical shell in the two configurations. The displacement vector is expressed as $\bm{u} = u(\rho)\bm{e}_{\rho}$. The covariant and contravariant vectors at the mid-surface in the two configurations, along with the reference and deformed normals, are specified by
\begin{eqnarray}
 &\textcolor{white}{=}&   {\bm{A}}_1=R\bm{e}_\theta, \quad {\bm{A}}_2=\bm{e}_z, \quad {\bm{A}}^1=\frac{1}{R}\bm{e}_\theta, \quad {\bm{A}}^2= \bm{e}_z, \nonumber \\
    &\textcolor{white}{=}&  {\bm{a}}_1=r\bm{e}_\theta, \quad {\bm{a}}_2=\bm{e}_z, \quad {\bm{a}}^1=\frac{1}{r}\bm{e}_\theta, \quad {\bm{a}}^2= \bm{e}_z,\nonumber \\
   &\textcolor{white}{=}&   \bm{n}=\bm{N}=\bm{e}_\rho.
\end{eqnarray}
The azimuthal unit vector is denoted by $\bm{e}_\theta$.
The components of the covariant metric tensors at the mid-surface in the reference and deformed configurations are:
\begin{eqnarray}
    \left[{A}_{\alpha \beta}\right]=\begin{bmatrix}
{R}^2 & 0 \\
0 & 1 
\end{bmatrix} \quad  \text{and} \quad
  \left[{a}_{\alpha \beta}\right]=\begin{bmatrix}
{r}^2 & 0 \\
0 & 1 
\end{bmatrix},
\end{eqnarray}
along with the determinant of the covariant metric tensors at the mid-surface given by
\begin{equation}
    A={R}^2, \quad \text{and} \quad a={r}^2.
\end{equation}
The incompressibility condition \eqref{eq:incomp_surf} results in
\begin{equation}
     J_0-1=0 \Rightarrow \lambda=\sqrt{\frac{A}{a}}=\lambda_\theta^{-1},
\end{equation}
with $\lambda_\theta=r/R$ as the azimuthal stretch. The  non-zero components of the curvature tensor at the mid-surface are 
\begin{equation}
    {B}_{1}^{\ 1}=-\frac{1}{R}, \quad \text{and} \quad {b}_{1}^{\ 1}=-\frac{1}{r}.
\end{equation}

In this case, the in-plane governing Equation \eqref{eq:strong1} is trivially satisfied due to axisymmetry of the deformation.
Upon substituting $\overline{p}=[p_\mathrm{t}+p_\mathrm{b}]/2$, $ \widetilde{p}=p_\mathrm{t}-p_\mathrm{b}$ and the expression for stress from Equation \eqref{eq: examples neo hooke stress}  in the out-of-plane balance Equation \eqref{eq:strong2}, one obtains
%
\begin{eqnarray}
    \dfrac {p_\mathrm{b}-p_\mathrm{t}}{\mu}=  \dfrac{\widetilde{T}\left[\lambda-1\right] \dfrac{p_t}{\mu} + \widetilde{T}\left[1-\lambda^4\right]}{\dfrac{\widetilde{T}}{2} \left[\lambda-1\right] + \lambda^{-1}}
\end{eqnarray}
where $\widetilde{T}=T/R$ is the dimensionless thickness of the cylindrical shell. 
Pressures at the inner and outer surfaces of the cylinder are denoted by $p_b$ and $p_t$, respectively. 
When the response of the system is derived from the perspective of considering the cylinder as a three dimensional hyperelastic body, the response expression is  given by \citep{HAUGHTON1979489}
\begin{eqnarray}
   \dfrac {p_\mathrm{b}-p_\mathrm{t}}{\mu} = 6\left[\lambda_\mathrm{b}^{-4}-\lambda_\mathrm{a}^{-4}\right]  - 2 \left[\ln{\lambda_\mathrm{b}}-\ln{\lambda_\mathrm{a}}\right],
\end{eqnarray}
where $\lambda_\mathrm{a}$ and $\lambda_\mathrm{b}$ are the stretches at the inner and outer surfaces of the infinite cylinder and are related by the incompressibility constraint
\begin{eqnarray}
    \lambda_\mathrm{b}^2-1={\left[\frac{R_\mathrm{out}-R_\mathrm{in}}{R_\mathrm{in}}\right]}^2 \left[\lambda_\mathrm{a}^2-1\right],
\end{eqnarray}
where $R_\mathrm{in}$ and $R_\mathrm{out}$ ane the radius at the inner and outer surfaces of the cylindrical shell.
Moreover, let $\lambda_\mathrm{c}$ represent the stretch at the mid-surface of the cylinder, which can be similarly connected to the stretches at the inner and outer surfaces from the incompressibility relation to eventually obtain a response function for the difference in pressures corresponding to $\lambda_\mathrm{c}$. These 3D relations allow for direct comparison with the shell system of equations.

The  comparison between the 3D and shell theory is presented in  Figure \ref{fig:res_2}(b) for various shell thickness values and external pressure values ($p_t/\mu$). 
As the pressure difference ($\Delta p = [p_b-p_t]/\mu$) between the inner and external shell surface increases, the stretch $\lambda_\theta$ monotonically increases until reaching a critical value  $\Delta p_\mathrm{c}$ corresponding to a limit point instability. 
At this juncture, minor changes in applied pressure lead to significant inflationary shifts. Similar limit point instabilities have been observed during inflation of thin hyperelastic shells and soft cylindrical cavities \citep{kiendl2015isogeometric, PhysRevLett.122.068003, MEHTA2022104031}. 
The critical limit point pressure decreases with decreasing shell thickness.
This example  demonstrates the distinction between considering the pressure on the top and bottom surfaces of the shell separately, as opposed to the common practice  of considering a pressure difference on the mid-surface. 
Varying the pressure on the external surface $p_t/\mu$ significantly alters the shell's response to the applied pressure difference $\Delta p$. 
Reduction in shell thickness brings these response curves closer together, as observed for $\widetilde{T}=1/30$ and $1/20$.
\section{Concluding Remarks}
\label{sec: conclusions}

This study rigorously derived the governing equations for the embedding aware large deformation of Kirchhoff-Love hyperelastic thin shells, using a variational approach and dimension reduction.
The general deformation map  within the shell was redefined to introduce terms for through-thickness stretch and the deformed normal.
Deformation function for the shell's mid-surface and the Lagrange multiplier associated with the incompressibility constraint were employed as the generalised set of solution variables for a two-field variational formulation, leading to the derivation of equilibrium equations for the shell.

The formulation preserves boundary conditions at the top and bottom surfaces of the shell thereby accounting for the fields in the embedding space.
In the current work, this manifests as a distinction between hydrostatic pressures applied to the top and bottom surfaces.
This  lays a foundation for multi-physics shell models of electro- and magnetoelasticity that must account for perturbations in the electromagnetic fields in the embedding media due to shell deformation. 
This enriched approach effectively models large deformations and limit point instabilities, as demonstrated by scenarios such as the stretching and squashing of a soft thin plate and the inflation of an infinitely long cylindrical hyperelastic shell.

This work provides a fresh perspective on shell system of equations, emphasising geometric precision and accuracy by capturing arbitrary geometric and constitutive nonlinearities.
The derived equations, rooted in a variational formulation, are well-prepared for future numerical implementation via the finite element method.

\section*{Acknowledgements}
This work was supported by the UK Engineering and Physical Sciences Research Council (EPSRC) grants EP/V030833/1 and EP/R008531/1, and a Royal Society grant IES/R1/201122.

\begin{appendices}
\renewcommand{\theequation}{A.\arabic{equation}}
\section{Geometry of a Kirchoff-Love thin shell}
\label{app: KL shell geometry}
\subsection{The natural basis at the mid-surface}
The covariant basis vectors for the mid-surface in the reference and deformed configurations, respectively, can be computed as
\begin{equation}
\bm{A}_\alpha = \frac{\partial {\bm{x}}}{\partial \theta^\alpha}, 
\quad \mathrm{and} \quad
\bm{a}_\alpha = \frac{\partial \bm{x}}{\partial \theta^\alpha}.
\end{equation}
Thus, the unit normal vectors in the two configurations are defined by
\begin{equation}
\bm{N} = \frac{{\bm{A}}_1 \times {\bm{A}}_2}{{A}^{1/2}}, \quad \text{and} \quad  \bm{n} = \frac{\bm{a}_1 \times \bm{a}_2}{{a^{1/2}}},
\label{eq:normal}
\end{equation}
where $A$ and $a$ are 
\begin{equation}
A=\norm{ {\bm{A}}_1 \times {\bm{A}}_2}^{2},\quad \text{and} \quad a = \norm{\bm{a}_1 \times \bm{a}_2}^{2}.
\end{equation}
Further, it can be shown that
\begin{equation}
  A= \mathrm{det}\left[{A}_{\alpha \beta}\right],\quad \text{and} \quad a = \mathrm{det}[{a}_{\alpha \beta}]. 
\end{equation}
The covariant components of the metric tensor for the mid-surface points $\bm{R}$ and $\bm{r}$ are respectively given by
\begin{equation}
{A}_{\alpha \beta} ={\bm{A}}_\alpha \cdot {\bm{A}}_\beta, \quad \text{and} \quad {a}_{\alpha \beta} = \bm{{a}}_{\alpha } \cdot \bm{{a}}_{ \beta}.
\end{equation}
Also, the contravariant metric tensor components for the mid-surface are 
\begin{equation}
{A}^{\alpha \gamma} {A}_{\gamma \beta} = \delta^{\alpha}_{\beta}, \quad \text{and} \quad {a}^{\alpha \gamma} {a}_{\gamma \beta} = \delta^{\alpha}_{\beta},
\label{eq:co_and_contra_metric}
\end{equation}
where $\delta^{\alpha}_{\beta}$ denotes the Kronecker Delta. Again,  ${\bm{A}}^{\alpha}$ and $\bm{a}^{\alpha}$ denotes the contravariant basis vectors for the mid-surface in the two configurations, and they are defined as 
\begin{equation}
{\bm{A}}^{\alpha} \cdot {\bm{A}}
_{\beta} = \delta^\alpha_\beta, \quad \text{and} \quad \bm{a}^{\alpha} \cdot \bm{a}_{\beta} = \delta^\alpha_\beta .
\label{eq:deform_tensor}
\end{equation} 
\subsection{The unit alternator and permutation symbol}
In general, for a surface tensor ${\bm{Q}}$, the surface inverse ${\bm{Q}}^{-1}$ defined from
\begin{equation}
    {\bm{Q}}^{-1}{\bm{Q}}={\bm{I}},
\end{equation}
with $\bm{I}={\bm{A}}^{\beta} \otimes {\bm{A}}_{\beta} = {\bm{A}}_{\beta} \otimes {\bm{A}}^{\beta} $ ($\bm{I}$ denotes  the projection onto the tangent plane of ${S}_{\mathrm{m}}$) has the contravariant components as
\begin{equation}
  {Q}_{\mathrm{inv}}^{\alpha \beta}=\frac{1}{Q} {e}^{\alpha \gamma} {Q}_{ \delta \gamma}{e}^{\beta \delta},
  \label{surf_inv}
\end{equation}
where $Q=\mathrm{det}\left[{Q}_{ \alpha \beta}\right]$, and  the so-called unit alternator given as  
\begin{equation}
    \left[e^{\alpha \gamma}\right]=\begin{bmatrix}
0 & 1 \\
-1 & 0 
\end{bmatrix}.
\end{equation}
Further, the permutation tensor is defined as
\begin{equation}
    \bm{E}={E}^{\alpha \beta} \ {\bm{A}}_{\alpha} \otimes {\bm{A}}_{\beta}= \frac{1}{{A}^{1/2}}{e}^{\alpha \beta} \ {\bm{A}}_{\alpha} \otimes {\bm{A}}_{\beta}, \quad \text{and} \quad  \bm{\varepsilon}={\varepsilon}^{\alpha \beta} \ \bm{a}_{\alpha} \otimes \bm{a}_{\beta}= \frac{1}{{a}^{1/2}}e^{\alpha \beta} \ \bm{a}_{\alpha} \otimes \bm{a}_{\beta},
    \label{eq:permutation_symbol_1}
\end{equation}
in the two configurations. In particular, Equation \eqref{surf_inv} yields
\begin{equation}
    {A}^{\alpha \beta}=\frac{1}{{A}} {e}^{\alpha \gamma} {A}_{\gamma \delta} {e}^{\beta \delta}, 
    \label{alternate_exp_metric_1}
\end{equation}
and similarly, 
\begin{equation}
     a^{\alpha \beta}=\frac{1}{a} e^{\alpha \gamma}a_{\gamma \delta}e^{\beta \delta}.
     \label{alternate_exp_metric_2}
\end{equation}
From the above, and using the relation, $e^{\gamma \alpha }e_{\gamma \beta}=\delta^\alpha_\beta$, we get  
\begin{equation}
  {A}^{\alpha \beta} {e}_{\beta \gamma} {A}= {e}^{\alpha \beta}{A}_{ \beta \gamma}\quad \text{and} \quad
   {a}^{\alpha \beta} {e}_{\beta \gamma}a= {e}^{\alpha \beta} {a}_{ \beta \gamma},
\end{equation}
which can be further used to rewrite the permutation tensors as
\begin{equation}
    \bm{E}={E}_{\alpha \beta}\ {\bm{A}}^{\alpha} \otimes {\bm{A}}^{\beta}= {A}^{1/2} {e}_{\alpha \beta} \ {\bm{A}}^{\alpha} \otimes {\bm{A}}^{\beta}, \quad 
    \text{and} 
    \quad  \bm{\varepsilon}={\varepsilon}_{\alpha \beta}\ \bm{a}^{\alpha} \otimes \bm{a}^{\beta}= {a}^{1/2}e_{\alpha \beta} \ \bm{a}^{\alpha} \otimes \bm{a}^{\beta}.
    \label{eq:permutation_symbol_2}
\end{equation}
Again, multiplying Equations \eqref{alternate_exp_metric_1} and \eqref{alternate_exp_metric_2} by ${A}_{\alpha \beta}$ and $a_{\alpha \beta}$, respectively, we obtain
\begin{equation}
   A=\frac{1}{2} {e}^{\alpha \gamma} {e}^{\beta \delta} {A}_{\alpha \beta} {A}_{\gamma \delta}, 
   \quad \text{and} \quad 
   a=\frac{1}{2} e^{\alpha \gamma} e^{\beta \delta} a_{\alpha \beta} a_{\gamma \delta}.
    \label{alternate_exp_det}  
\end{equation}
From the above, we can write
\begin{equation}
  {A}_{,\zeta}=  {e}^{\alpha \gamma} {e}^{\beta \delta} {A}_{\alpha \beta, \zeta} {A}_{\gamma \delta}, \quad \text{and} \quad 
  a_{,\zeta}= e^{\alpha \gamma} e^{\beta \delta} a_{\alpha \beta, \zeta} a_{\gamma \delta},
\end{equation}
which can be further rewritten as
\begin{equation}
      {A}_{,\zeta}= {A} {A}^{\alpha \beta}  {A}_{\alpha \beta, \zeta}, \quad \text{and} \quad 
  a_{,\zeta}= a a^{\alpha \beta}  a_{\alpha \beta, \zeta} ,
  \label{det_metric_parametric_der}
\end{equation}
by using Equations \eqref{alternate_exp_metric_1} and \eqref{alternate_exp_metric_2}. Now,
\begin{equation}
 {A}_{\alpha \beta, \zeta}={\bm{A}}
_{\alpha, \zeta}\cdot {\bm{A}}
_{\beta} + {\bm{A}}
_{\alpha}\cdot {\bm{A}}
_{\beta, \zeta} , \quad\text{and} \quad 
 {a}_{\alpha \beta, \zeta}={\bm{a}}
_{\alpha, \zeta}\cdot{\bm{a}}
_{\beta} + {\bm{a}}
_{\alpha}\cdot{\bm{a}}
_{\beta, \zeta}.
\end{equation}
Therefore, 
\begin{equation}
   {A}_{,\zeta}= 2{A} {\Gamma}^\alpha_{\alpha \zeta}, \quad \text{and} \quad {a}_{,\zeta}= 2{a} {\gamma}^\alpha_{\alpha \zeta},
   \label{parametric_der_christoffel}
\end{equation}
with the  Christoffel symbols of the second kind in the two configurations defined by 
\begin{equation}
  {\Gamma}^\alpha_{\zeta \gamma}= {\bm{A}}^{\alpha} \cdot {\bm{A}}_{\zeta, \gamma}, \quad \text{and} \quad  {\gamma}^\alpha_{\zeta \gamma}= {\bm{a}}^{\alpha} \cdot {\bm{a}}_{\zeta, \gamma}.
\end{equation}
\subsection{The natural basis at a shell-point}
\label{app: natural basis}
A point $\bm{x} \in \mathcal{B}$ can be written as 
\begin{equation}
\bm{x}=\bm{r}+\eta \bm{d}, 
\end{equation} 
where $\bm{d}=\lambda \bm{n}$ and $\lambda = \dfrac{t}{T}$. 
The covariant basis vectors at a point  $\bm{x}$  in the shell are 
\begin{eqnarray}
\bm{g}_{\alpha}&=&\frac{\partial \bm{x}}{\partial \theta^{\alpha}}, \nonumber \\
&=&\frac{\partial \bm{r}}{\partial \theta^{\alpha}} +\eta \bm{n} \frac{\partial \lambda}{\partial \theta^{\alpha}}+\eta \lambda \frac{\partial \bm{n}}{\partial \bm{r}} \frac{\partial \bm{r}}{\partial \theta^{\alpha}}, \nonumber \\
&=& \bm{\mu}\bm{a}_{\alpha}.
\end{eqnarray}
In the above,  neglecting the variation of the through-thickness stretch along the mid-surface of the shell-structure \citep{ kiendl2015isogeometric, liu2022computational}, 
we arrive at
\begin{equation}
    \bm{\mu}=\bm{i}-\eta \lambda \bm{\kappa},
    \label{eq:shiftor2}
\end{equation}
where  $\bm{i}=  \bm{a}^{\beta} \otimes \bm{a}_{\beta}=\bm{a}_{\beta} \otimes \bm{a}^{\beta}$ denotes  the projection onto the tangent plane of $s_\mathrm{m}$, the  deformed counterpart of ${S}_\mathrm{m}$. Also,
\begin{equation}
    \bm{\kappa}=-\dfrac{\partial \bm{n}}{\partial \bm{r}}
    =-{\bm{n}}_{, \beta} \otimes {\bm{a}}^{\beta}.
\end{equation}
The covariant basis vectors at a point  $\bm{X}$  in the shell are given by 
\begin{eqnarray}
{\bm{G}}_{\alpha}&=&\frac{\partial \bm{X}}{\partial \theta^{\alpha}}, \nonumber \\
&=&\frac{\partial \bm{R}}{\partial \theta^{\alpha}} + \eta \frac{\partial \bm{N}}{\partial \bm{R}} \frac{\partial \bm{R}}{\partial \theta^{\alpha}}, \nonumber \\
&=& \bm{M}{\bm{A}}_{\alpha},
\end{eqnarray}
where
\begin{equation}
    \bm{M}=\bm{I}-\eta \bm{K},
    \label{eq:shiftor1}
\end{equation}
with  
\begin{equation}
  \bm{K} =-\dfrac{\partial \bm{N}}{\partial \bm{R}}=-{\bm{N}}_{, \beta} \otimes {\bm{a}}^{\beta}.
\end{equation}
Again, ${\bm{N}}_{, \beta}$ and ${\bm{n}}_{, \beta}$ appears in the formula of Weingarten as
\begin{equation}
    {\bm{N}}_{, \beta}=-{B}_\beta^{ \ \gamma} \ {\bm{A}}_{\gamma}, \quad \text{and} \quad {\bm{n}}_{, \beta}=-{{b}}_\beta^{ \ \gamma} \ {\bm{a}}_{\gamma},
\end{equation}
with the surface curvature tensors in the two configurations defined as
\begin{equation}
 \bm{B}
={B}_{\beta \delta} \ {\bm{A}}^{\beta} \otimes {\bm{A}}^{\delta}\quad \text{and} \quad {\bm{b}}
={b}_{\beta \delta} \ {\bm{a}}^{\beta} \otimes {\bm{a}}^{\delta},   
\end{equation}
where
\begin{equation}
  {B}_{\beta \delta}=\bm{N} \cdot {\bm{A}}_{\beta , \delta},  \quad \text{and} \quad {b}_{\beta \delta}={\bm{n}} \cdot {\bm{a}}_{\beta , \delta},
\end{equation}
and further,
\begin{equation}
 {{B}_\beta}^{\gamma}={B}_{\beta\delta}{A}^{\delta \gamma},\quad \text{and} \quad {{b}_\beta}^{\gamma}={b}_{\beta \delta}{a}^{\delta \gamma}. 
\end{equation}
Therefore,
\begin{equation}
  \bm{K} ={B}_\beta^{ \ \gamma} \ {\bm{A}}_{\gamma}\otimes {\bm{A}}^{\beta}, \quad \text{and} \quad  \bm{\kappa}={{b}}_\beta^{ \ \gamma} \ {\bm{a}}_{\gamma}\otimes {\bm{a}}^{\beta}.
\end{equation}
For a point in the shell, the components of the covariant and contravariant metric tensors in the reference configuration are defined by
\begin{equation}
{G}_{\alpha \beta} = {\bm{G}}_{\alpha} \cdot {\bm{G}}_{\beta}\quad \text{and} \quad {G}^{\alpha \beta} = {\bm{G}}^{\alpha} \cdot {\bm{G}}^{\beta},
\end{equation}
with the deformed counterparts as
\begin{equation}
    g_{\alpha \beta} = \bm{g}_{\alpha} \cdot \bm{g}_{\beta} \quad \text{and} \quad g^{\alpha \beta} = \bm{g}^{\alpha} \cdot \bm{g}^{\beta},
    \label{eq:contra_metric}
\end{equation}
where ${\bm{G}}^{\alpha}$ and $\bm{g}^{\alpha}$ denotes the contravariant basis vectors in the two configurations defined by
\begin{equation}
{\bm{G}}^{\alpha} \cdot {\bm{G}}
_{\beta} = \delta^\alpha_\beta, \quad \text{and} \quad \bm{g}^{\alpha} \cdot \bm{g}_{\beta} = \delta^\alpha_\beta  .
\label{eq:deform_tensor_1}
\end{equation}
It can be shown that
\begin{equation}
    {\bm{G}}^{\alpha}={\bm{M}}^{-\mathrm{T}} {\bm{A}}^{\alpha},
     \label{def_grad_deriv_2}
\end{equation}
and  ${\bm{M}}^{-1}$ can be expanded as
\begin{eqnarray}
    {\bm{M}}^{-1}
    &=&{M}^{-1\gamma}_{0_\beta}\ {\bm{A}}_{\gamma} \otimes {\bm{A}}^{\beta} + 
    \eta {M}^{-1  \gamma}_{1_\beta}\ {\bm{A}}_{\gamma} \otimes {\bm{A}}^{\beta} + 
    \eta^2 {M}^{-1  \gamma}_{2_\beta}\ {\bm{A}}_{\gamma} \otimes {\bm{A}}^{\beta}+ \bm{\mathcal{O}}(\eta^3),
    \label{def_grad_deriv_1}
\end{eqnarray}
with
\begin{equation}
{M}^{-1\gamma}_{0_\beta}= \delta^\gamma_\beta, \quad
{M}^{-1  \gamma}_{1_\beta}= {B}_\beta^{ \ \gamma}, \quad \text{and} \quad
{M}^{-1  \gamma}_{2_\beta}= {B}_\delta^{ \ \gamma}  {B}_\beta^{ \ \delta}. 
\end{equation}
Similarly,
\begin{equation}
    \bm{g}^{\alpha}=\bm{\mu}^{-\mathrm{T}}\bm{a}^{\alpha},
     \label{def_grad_deriv_3}
\end{equation}
and  $\bm{\mu}^{-1}$ can be expanded as
\begin{equation}
    \bm{\mu}^{-1}
    ={\mu}^{-1  \gamma}_{0_\beta}\ {\bm{a}}_{\gamma} \otimes {\bm{a}}^{\beta} + 
    \eta {\mu}^{-1  \gamma}_{1_\beta}\ {\bm{a}}_{\gamma} \otimes {\bm{a}}^{\beta} + 
    \eta^2 {\mu}^{-1  \gamma}_{2_\beta}\ {\bm{a}}_{\gamma} \otimes {\bm{a}}^{\beta}+ \bm{\mathcal{O}}(\eta^3),
    \label{def_grad_deriv_4}
\end{equation}
with
\begin{equation}
{\mu}^{-1  \gamma}_{0_\beta}= \delta^\gamma_\beta,\quad
{\mu}^{-1  \gamma}_{1_\beta}= \lambda{b}_\beta^{ \ \gamma}, \quad \text{and} \quad
{\mu}^{-1  \gamma}_{2_\beta}= \lambda^2{b}_\delta^{ \ \gamma} \ {b}_\beta^{ \ \delta}. 
\end{equation}
\subsubsection{The volume and surface elements}
The volume element in the reference configuration can be expressed as
\begin{eqnarray}
dV&=&\left[{\bm{G}}_1 \times {\bm{G}}_2\right] \cdot \bm{N} \ d\theta^1 d\theta^2 d\eta, \nonumber\\
&=&  \ \left[{\bm{A}}_1 \times {\bm{A}}_2\right] \cdot \bm{N}  M d\theta^1 d\theta^2 d\eta, \nonumber\\
&=&dS d\eta,
\label{diff_vol}
\end{eqnarray}
where the undeformed elemental area $dS$ is given by 
\begin{equation}
dS=M d{S}_\mathrm{m} ,
\label{eq:undeformed_area_arb}
\end{equation}
with $d{S}_\mathrm{m}$ as the area element on ${S}_\mathrm{m}$ written as
\begin{equation}
    dS_{\mathrm{m}}={A}^{1/2} dP,
    \label{eq:undeformed_area_mid}
\end{equation}
and the area element for the convected coordinates is $ dP=d\theta^1 d\theta^2$.
Also,
\begin{eqnarray}
M&=&\mathrm{det}{\bm{M}}, \nonumber \\
&=&1-2\eta H+\eta^2 K,
\label{eq:shift_ref}
\end{eqnarray}
where $H$ and $K$ are the mean and Gaussian curvatures of the undeformed mid-surface, respectively, and are expressed as
\begin{equation}
    H=\frac{1}{2}\mathrm{tr}{\bm{K}}=\frac{1}{2}\frac{\partial{\bm{N}}}{\partial {\bm{R}}}:{\bm{I}}=\frac{1}{2}{B}_{ \alpha \beta}{A}^{\alpha \beta}=\frac{1}{2}{B}_\alpha^{\alpha},
\end{equation}
and
\begin{equation}
    K=\mathrm{det}{\bm{K}}
    =\mathrm{det}\left[{B}_\alpha^{ \ \beta}\right]=\mathrm{det}\left[{B}_{ \alpha \gamma}{A}^{\gamma \beta}\right]=\frac{B}{A},
\end{equation}
with $B=\mathrm{det}\left[{B}_{ \alpha \beta}\right]$.
Further, an elemental area in the deformed configuration is given by
\begin{equation}
    ds=\mu {\hat{a}}^{1/2} dS_\mathrm{m},
\end{equation}
with 
the surface stretch $\hat{a}=\dfrac{a}{A}$, 
and
\begin{equation}
    \mu=1-2\eta \lambda h+\eta^2 \lambda^2 \kappa,
\end{equation}
where the mean and Gaussian curvatures of the deformed mid-surface are
\begin{equation}
     h=\frac{1}{2}{b}_\alpha^{\alpha},\quad \text{and} \quad \kappa=\frac{b}{a},
\end{equation}
with $b=\mathrm{det}[b_{ \alpha \beta}]$.
The boundaries, $\partial{\mathcal{B}}_{\mathrm{0}}$ and $\partial \mathcal{B}$ can be written as
   $ \partial{\mathcal{B}}_{\mathrm{0}}= S_\mathrm{t} \cup   S_\mathrm{b}  \cup   S_\mathrm{l}$, and $\partial \mathcal{B}= s_\mathrm{t} \cup  s_\mathrm{b} \cup  s_\mathrm{l}$,
where, the subscripts, t, b, and l, represents the top, bottom, and  lateral surfaces in the two configurations, and the top surface is the side of the boundary that is reached along the unit outward normal vector. Therefore,
\begin{equation}
   d S_\mathrm{t}={M}\Big|_{\eta=T/2}
   dS_\mathrm{m}, \quad \text{and} \quad  d S_\mathrm{b}={M}\Big|_{\eta=-T/2}
   d{S}_{\mathrm{m}}.
   \label{eq:ref_top_bot_to_mid}
\end{equation}
Also,
\begin{equation}
   d s_\mathrm{t}=\mu\Big|_{\eta=T/2}
   {\hat{a}}^{1/2} d S_\mathrm{m}, \quad \text{and} \quad  d s_\mathrm{b}=\mu\Big|_{\eta=-T/2}
   {\hat{a}}^{1/2} dS_\mathrm{m}.
   \label{eq:def_surface_top_bottom}
\end{equation}
If the bounding curve  $C_\mathrm{m}$ of  the mid-surface $S_\mathrm{m}$  is characterized by the arc-length parameter $l$, then the infinitesimal length $dl$ between two points ${\bm{R}}(\theta^1, \theta^2)$ and ${\bm{R}}(\theta^1 + d\theta^1, \theta^2+d\theta^2)$ is given by
\begin{eqnarray}
dl&=&\norm{{\bm{R}}(\theta^1 + d\theta^1, \theta^2+d\theta^2)-{\bm{R}}(\theta^1, \theta^2)}, \nonumber\\
&=&\norm{{\bm{R}}(\theta^1, \theta^2) + \frac{\partial \bm{R}}{\partial \theta^\alpha} d\theta^\alpha - {\bm{R}}(\theta^1, \theta^2)}, \nonumber \\
&=&\norm{{\bm{A}}_\alpha d\theta^\alpha}, \nonumber \\
&=& \sqrt{{\bm{A}}_\alpha d\theta^\alpha \cdot {\bm{A}}_\beta d\theta^\beta}, \nonumber \\
&=& \sqrt{A_{\alpha \beta}d\theta^\alpha d\theta^\beta}.
\label{eq:dl}
\end{eqnarray}
The tangent vector at a point $\bm{R}$ on $C_\mathrm{m}$ is defined as 
\begin{equation}
    \bm{\tau}=\frac{d\bm{R}}{dl}=\frac{\partial \bm{R}}{\partial\theta^\beta}\frac{d\theta^\beta}{dl}={\bm{A}}_\beta \frac{d\theta^\beta}{dl},
\end{equation}
and using Equation \eqref{eq:dl}, we get
\begin{eqnarray}
  \bm{\tau} \cdot \bm{\tau}&=&{\bm{A}}_\alpha \cdot {\bm{A}}_\beta \frac{d\theta^\alpha}{dl} \frac{d\theta^\beta}{dl}, \nonumber \\
  &=&{A}_{\alpha \beta} \frac{d\theta^\alpha}{dl} \frac{d\theta^\beta}{dl}, \nonumber \\
  &=&1,  
\end{eqnarray}
implying that $\bm{\tau}$ is a unit tangent vector. Further,  we define 
\begin{equation}
    \bm{\nu}=\bm{E}\bm{\tau}={E}^{\alpha \beta}{A}_{\beta \gamma}\frac{d\theta^\gamma}{dl} {\bm{A}}_\alpha={E}_{\eta \delta} \frac{d \theta^\delta}{dl} {\bm{A}}^\eta,
    \label{eq:in-plane_normal}
\end{equation}
such that,
\begin{eqnarray}
     \bm{\nu} \cdot \bm{\tau}&=&{E}_{\alpha \beta} \frac{d \theta^\alpha}{dl} \frac{d \theta^\beta}{dl},\nonumber \\
     &=&\frac{1}{2}\left[{E}_{\alpha \beta}+{E}_{\beta \alpha}\right]\frac{d \theta^\alpha}{dl} \frac{d \theta^\beta}{dl}, \nonumber \\
     &=&0.
\end{eqnarray}
Again,
\begin{eqnarray}
  \bm{\nu} \cdot  \bm{\nu} &=& {E}^{\alpha \beta}{A}_{\beta \gamma}\frac{d\theta^\gamma}{dl} {\bm{A}}_\alpha \cdot {E}_{\eta \delta} \frac{d \theta^\delta}{dl} {\bm{A}}^\eta , \nonumber \\
  &=&{E}^{\alpha \beta}{E}_{\eta \delta}\delta^{\eta}_\alpha {A}_{\beta \gamma} \frac{d\theta^\gamma}{dl}\frac{d \theta^\delta}{dl},
\end{eqnarray}
and following the relation,
   $ {E}^{\alpha \beta}{E}_{\eta \delta}\delta^{\eta}_\alpha=\left[\delta^{\alpha}_\eta\delta^{\beta}_\delta-\delta^{\alpha}_\delta\delta^{\beta}_\eta\right]\delta^{\eta}_\alpha
    =\delta^{\beta}_\delta$,
we get
\begin{eqnarray}
  {\bm{\nu}} \cdot \bm{\nu}&=&\delta^{\beta}_\delta {A}_{\beta \gamma} \frac{d\theta^\gamma}{dl}\frac{d \theta^\delta}{dl}, \nonumber \\
  &=&{A}_{\delta \gamma} \frac{d\theta^\gamma}{dl}\frac{d \theta^\delta}{dl}, \nonumber \\
  &=& \bm{\tau} \cdot \bm{\tau},\nonumber \\
  &=&1,  
\end{eqnarray}
implying that $\bm{\nu}$ is the in-plane unit normal to $\bm{\tau}$ on $\mathcal{C}_\mathrm{m}$, and
\begin{equation}
    \bm{\nu}=\bm{\tau}\times \bm{N}.
\end{equation}
An elemental area, $d {S}_\mathrm{l}$, at a point $\bm{X}$ on the lateral surface is given by
\begin{eqnarray}
d S_\mathrm{l}&=&\norm{\frac{\partial \bm{X}}{\partial l}\times \frac{\partial \bm{X}}{\partial \eta}}dl \, d\eta, \nonumber \\
&=&\norm{\bm{M}\bm{\tau} \times \bm{N}}dl \, d\eta, \nonumber \\
&=&c\norm{{\bm{\tau}}_\mathrm{l}\times \bm{\bm{N}}}dl \, d\eta, \nonumber \\
&=&c\, dl \, d\eta,
\label{eq:lateral_to_mid}
\end{eqnarray}
with
\begin{equation}
    c=\norm{\bm{M}\bm{\tau}}={\left[1-2\eta \bm{K} \bm{\tau} \cdot \bm{\tau}+\eta^2 \bm{K}\bm{\tau} \cdot \bm{K}\bm{\tau}\right]}^{1/2}, \quad \text{and} \quad \bm{\tau}_\mathrm{l}=\frac{\bm{M}\bm{\tau}}{c},
\end{equation}
where ${\bm{\tau}}_\mathrm{l}$ is the unit tangent vector at a point on the lateral surface on the bounding curve $\mathcal{C}$ of the surface $S$. The in-plane unit normal to ${\bm{\tau}}_\mathrm{l}$ is given by
\begin{equation}
    {\bm{\nu}}_\mathrm{l}={\bm{\tau}}_{\mathrm{l}}\times\bm{\bm{N}}.
\end{equation}
The above can be written as
\begin{equation}
     c{\bm{\nu}}_\mathrm{l} =\left[\bm{I}+\eta\left[\bm{K}-2H\bm{I}\right]\right]\bm{\nu},
     \label{eq:lateral_nor_to_mid}
\end{equation}
by using the relation,
\begin{equation}
     \bm{K}\bm{\tau}\times \bm{N}=\left[2H\bm{I}- 
     \bm{K}\right]\bm{\nu}.
\end{equation}
\end{appendices}
\begin{appendices}
\renewcommand{\theequation}{B.\arabic{equation}}
\section{Application of Green's Theorem at the mid-surface of the shell}
\label{app: Green}
For a scalar ${T}^\alpha$, consider the following integral:
\begin{equation}
\int\limits_{ P}\left[{A}^{1/2}{T}^\alpha\right]_{, \alpha}dP,
\end{equation}
which can be rewritten by applying the Green's theorem as
\begin{eqnarray}
  \int\limits_{ P}\left[{A}^{1/2}{T}^\alpha\right]_{, \alpha}dP &=& \int\limits_{ P}\left[\left[{A}^{1/2}{T}^1\right]_{, 1}+\left[{A}^{1/2}{T}^2\right]_{, 2}\right]dP, \nonumber \\
  &=& \int\limits_{\mathcal{C}_\mathrm{p}} {A}^{1/2} {e}_{\alpha \beta}{T}^\alpha d\theta^\beta,
\end{eqnarray}
where $\mathcal{C}_\mathrm{p}$ is the boundary of the parametric domain $P$, and the above boundary integral can be simplified as
\begin{eqnarray}
  \int\limits_{\mathcal{C}_\mathrm{p}} {A}^{1/2} {e}_{\alpha \beta}{T}^\alpha d\theta^\beta
  &=&\int\limits_{\mathcal{C}_\mathrm{p}} {E}_{\alpha \beta}{T}^\alpha d\theta^\beta, \nonumber \\
  &=&\int\limits_{\mathcal{C}_\mathrm{m}}\left[{E}_{\alpha \beta}\frac{d\theta^\beta}{dl}\right]{T}^\alpha dl, 
\end{eqnarray}
which on using Equation \ref{eq:in-plane_normal} can be further wriiten as
\begin{equation}
 \int\limits_{\mathcal{C}_\mathrm{p}} {A}^{1/2} {e}_{\alpha \beta}{T}^\alpha d\theta^\beta=\int\limits_{\mathcal{C}_\mathrm{m}}{T}^\alpha {\nu}_\alpha dl.   
\end{equation}
Therefore, we relate the integral over the parametric domain to the line-integral along the boundary of the curved mid-surface.
\end{appendices}
\begin{appendices}
\renewcommand{\theequation}{C.\arabic{equation}}
\section{Variation of some relevant quantities}
Here, we list the first variation of key kinematic variables (essential for the calculations in Sec. 4),
for example,
\begin{equation}
    \delta\bm{F}=\delta\frac{\partial \bm{\chi}}{\partial \bm{X}}
    =\frac{\partial \delta\bm{\chi}}{\partial \bm{X}},
\end{equation}
and following
\begin{equation}
    \bm{F}\bm{F}^{-1}=\mathbold{\mathbb{1}},
\end{equation}
we obtain
\begin{equation}
    \delta \bm{F}^{-1}=-\bm{F}^{-1} \delta \bm{F}\bm{F}^{-1}. 
\end{equation}
Also, 
\begin{equation}
    \delta J=J\bm{F}^{-T}:\delta\bm{F}.
    \label{eq:var_jac}
\end{equation}
Again, 
\begin{eqnarray}
    \delta\bm{\chi}&=&\delta \bm{r}+\eta \delta \bm{d}, \nonumber \\
    &=&\delta \bm{r}+\eta \delta\lambda  \bm{n} + \eta  \lambda  \delta\bm{n}, 
    \label{eq:var_chi_1}
\end{eqnarray}
where $\delta \bm{n}$ can be obtained by using the relations, $\bm{n}\cdot\bm{n}=1$ and $\bm{a}_{\alpha}\cdot\bm{n}=0$ as
\begin{equation}
    \delta \bm{n}=-\left[\bm{a}^\alpha \otimes \bm{n}\right]\delta \bm{a}_\alpha=-\bm{a}^\alpha \left[\bm{n} \cdot \delta \bm{a}_\alpha\right],
    \label{eq:var_def_nor}
\end{equation}
with
\begin{equation}
    \delta \bm{a}_\alpha=\delta \frac{\partial \bm{r}}{\partial \theta^\alpha}=\left[\delta \bm{r}\right]_{, \alpha},
    \label{eq: delta a lpha}
\end{equation}
and moreover, from Equation \eqref{eq:through_thickness} follows
\begin{equation}
    \delta \lambda= -\frac{\lambda}{2} a^{-1}\delta a,
\end{equation}
where from Equation \eqref{alternate_exp_det},
\begin{equation}
    \delta a=a a^{\alpha \beta}\delta a_{\alpha \beta},
\end{equation}
 which can be rewritten by using,
\begin{equation}
    \delta a_{\alpha \beta}=\delta {\bm{a}}_{\alpha} \cdot  {\bm{a}}_{\beta} +  {\bm{a}}_{\alpha} \cdot \delta {\bm{a}}_{\beta},
\end{equation}
as
\begin{equation}
    \delta a= 2a \bm{a}^{\alpha}\cdot\delta {\bm{a}}_{\alpha}.
\end{equation}
Therefore, the variation in the through-thickness stretch can be rewritten as
\begin{equation}
    \delta \lambda=
    -\lambda \bm{a}^{\alpha}\cdot\delta {\bm{a}}_{\alpha}. 
    \label{eq:var_lambda}
\end{equation}
Also, 
\begin{equation}
   \delta p=\delta p_0+\eta \delta p_1
   +\mathcal{O}\left(\eta^2\right),
\end{equation}
and $p=p_0+\eta p_1
+\mathcal{O}\left(\eta^2\right)$.
\end{appendices}

\bibliographystyle{author-year-prashant}
\bibliography{ref1} 
\end{document}